\documentclass[10pt,twocolumn]{IEEEtran}
\IEEEoverridecommandlockouts
\usepackage{cite}
\usepackage{amsmath,amssymb,amsfonts, acronym}
\usepackage[ruled,vlined]{algorithm2e}
\usepackage{psfrag}
\usepackage{soul}
\usepackage{graphicx}
\usepackage{textcomp}
\usepackage[table]{xcolor}
\usepackage{bbm}
\usepackage{caption}
\usepackage{subcaption}
\usepackage{gensymb}

\usepackage{hyperref}
\usepackage{tikz}
\usepackage{pgfplots}
\usepackage{xparse}

\usepackage{comment}

\usepackage{bbm}

\newcommand{\red}[1]{\textcolor{red}{ {#1}}}

\def\BibTeX{{\rm B\kern-.05em{\sc i\kern-.025em b}\kern-.08em
 T\kern-.1667em\lower.7ex\hbox{E}\kern-.125emX}}

\definecolor{silver}{rgb}{0.95, 0.95, 0.95}

\acrodef{5G}{fifth generation}
\acrodef{6G}{sixth generation}
\acrodef{AOA}{angle-of-arrival}
\acrodef{AOD}{angle-of-departure}
\acrodef{AI}{Artificial Intelligence}
\acrodef{UAV-I}{{\em UAV intelligence}}
\acrodef{AI/ML}{artificial intelligence/machine learning}
\acrodef{CNPC}{control and non-payload communication}
\acrodef{A2A}{agent-to-agent}
\acrodef{AWGN}{additive white Gaussian noise}
\acrodef{AO}{alternating optimization}
\acrodef{BS}{base station}
\acrodef{BCD}{block coordinate descent}
\acrodef{BD}{Block Diagonalization}
\acrodef{CRLB}{Cram\'er-Rao lower bound}
\acrodef{CRB}{Cram\'er-Rao bound}
\acrodef{CDF}{cumulative distribution function}
\acrodef{cdf}{cumulative distribution function}
\acrodef{CDR}{correct detection rate}
\acrodef{CIR}{channel impulse response}
\acrodef{CRAS}{connected robotics and autonomous systems}
\acrodef{CSI}{channel state information}
\acrodef{DEC-POMDP}{decentralized partially observable Markov decision process}
\acrodef{D2D}{device-to-device}
\acrodef{EKF}{extended Kalman filter}
\acrodef{EIRP}{effective isotropic radiated power}
\acrodef{KF}{Kalman filter}
\acrodef{EM}{electromagnetic}
\acrodef{FAR}{false alarm rate}
\acrodef{FIM}{Fisher Information Matrix}
\acrodef{FL}{federated learning}
\acrodef{FMCW}{frequency modulated continuous wave}
\acrodef{GLRT}{generalized likelihood ratio test}
\acrodef{GP}{Gaussian process}
\acrodef{GPI}{generalized policy iteration}
\acrodef{IoT}{Internet-of-Things}
\acrodef{IS}{image similarity}
\acrodef{IRS}{intelligent reflecting surfaces}
\acrodef{KPI}{key performance indicator}
\acrodef{KKT}{Karush Kuhn Tucker }
\acrodef{LLRT}{log-likelihood ratio test}
\acrodef{LOS}{line-of-sight}
 \acrodef{MAC}{medium access control}
\acrodef{MAP}{maximum a-posteriori probability}
\acrodef{MEC}{multi-access edge computing}
\acrodef{MAB}{multi-armed bandit}
\acrodef{mMTC}{massive machine type communication}
\acrodef{eMBB}{enhanced mobile broadband}

\acrodef{MMSE}{minimum mean squared error }
\acrodef{MARL}{multi-agent reinforcement learning}
\acrodef{MDP}{Markov decision process}
\acrodef{MIMO}{multiple-input multiple-output}
\acrodef{ML}{Machine Learning}
\acrodef{MLE}{maximum likelihood estimator}
\acrodef{mm-wave}{millimeter-wave}
\acrodef{MSE}{mean squared error}
\acrodef{NOMA}{non-orthogonal multiple access}
\acrodef{NLOS}{non-line-of-sight}
\acrodef{NF}{Noise Factor}
\acrodef{NSPD}{Noise Power Spectral Density}
\acrodef{OG}{occupancy grid}
\acrodef{OFDM}{Orthogonal Frequency Division Multiplexing}
\acrodef{PF}{particle filtering}
\acrodef{pdf}{probability density function}
\acrodef{PFA}{probability of false alarm}
\acrodef{PL}{packet loss}
\acrodef{POMDP}{partially observable Markov decision process}
\acrodef{PER}{packet error rate}
\acrodef{PEB}{position error bound}
\acrodef{RIS}{reconfigurable intelligent surface}
\acrodef{RCS}{radar cross section}
\acrodef{RMSE}{root mean square error}
\acrodef{RFID}{radiofrequency identification}
\acrodef{RL}{reinforcement learning}
\acrodef{ROC}{receiver operating characteristics}
\acrodef{RR}{reading range}
\acrodef{RRCS}{root-radar cross section}
\acrodef{RSS}{received signal strength}
\acrodef{RV}{random variable}
\acrodef{SLAM}{simultaneous localization and mapping}
\acrodef{SNR}{signal-to-noise ratio}
\acrodef{SIR}{sequential importance resampling} 
\acrodef{SRE}{smart radio environment} 
\acrodef{SISO}{single-input single-output} 
\acrodef{SVD}{Singular Value Decomposition}
\acrodef{TD}{temporal-difference}
\acrodef{TOA}{time-of-arrival}
\acrodef{UE}{user equipment}
\acrodef{THz}{Terahertz}
\acrodef{UAV}{unmanned aerial vehicle}
\acrodef{U2U}{UAV-to-UAV}
\acrodef{UKF}{unscented Kalman filter}
\acrodef{UCB}{upper confidence bound}
\acrodef{URA}{uniform rectangular array}
\acrodef{ULA}{uniform linear array}
\acrodef{URLLC}{ultrareliable and low-latency communication}
\acrodef{UWB}{ultra wideband}
\acrodef{VLC}{visible light communication}

\acrodef{wMMSE}{weighted minimum mean square error}

\newcommand{\Nrx}{{N}_{\text{RX}}}
\newcommand{\Ntx}{{N}_{\text{TX}}}
\newcommand{\tempo}{\mathsf{t}}
\newcommand{\tempop}{\mathsf{t+1}}
\newcommand{\tempom}{\mathsf{t-1}}
\newcommand{\tx}{i}
\newcommand{\rx}{r}
\newcommand{\rr}{\tilde{r}}
\newcommand{\risk}{k}

\newcommand{\riselem}{p}

\newcommand{\prx}{\mathbf{p}_{\text{RX}}}
\newcommand{\pk}{\mathbf{p}_{\risk}}
\newcommand{\pRX}{\mathbf{p}_{\text{RX}}}
\newcommand{\pRXt}{\mathbf{p}_{\tempo}}
\newcommand{\pRXtp}{\mathbf{p}_{\tempop}}
\newcommand{\ptx}{\mathbf{p}_{\text{TX}}}
\newcommand{\prxr}{\mathbf{p}_{\text{RX},\rx}}
\newcommand{\ptxt}{\mathbf{p}_{\text{TX},\tx}}
\newcommand{\pkp}{\mathbf{p}_{\risk,\riselem}}
\newcommand{\pt}{\mathbf{p}_{\tx}}
\newcommand{\pr}{\mathbf{p}_{\rx}}
\newcommand{\dTXk}{d_{\risk, \text{TX}}}
\newcommand{\dTXRX}{d_{\text{RX}, \text{TX}}}
\newcommand{\dtr}{d_{\rx,\tx}}
\newcommand{\dktpr}{d_{\risk, \rx, \riselem, \tx}}
\newcommand{\dRXk}{d_{\text{RX}, \risk}}
\newcommand{\dkpr}{d_{\risk, \rx, \riselem}}
\newcommand{\dktp}{d_{\risk, \riselem, \tx}}

\newcommand{\dr}{\mathsf{d}_{\rx}}
\newcommand{\thetar}{\theta_{\rx}}
\newcommand{\phir}{\phi_{\rx}}
\newcommand{\dt}{\mathsf{d}_{\tx}}
\newcommand{\thetat}{\theta_{\tx}}
\newcommand{\phit}{\phi_{\tx}}
\newcommand{\dkp}{\mathsf{d}_{\risk, \riselem}}
\newcommand{\phikp}{\phi_{ \risk, \riselem}}
\newcommand{\thetakp}{\theta_{\risk, \riselem}}

\newcommand{\Thetaktp}{\Theta_{\risk, \riselem, \tx}}
\newcommand{\Thetakpr}{\Theta_{\risk, \rx, \riselem}}

\newcommand{\thetaktp}{\theta_{\risk, \riselem, \tx}}

\newcommand{\phiktp}{\phi_{\risk, \riselem, \tx}}

\newcommand{\ftk}{f_{\tx,\risk}}
\newcommand{\htr}{h_{\rx,\tx}}
\newcommand{\bkpr}{b_{\risk,\rx, \riselem} }
\newcommand{\gktp}{g_{\risk,\riselem, \tx }}
\newcommand{\rhoktp}{\rho_{\risk,\riselem,\tx}}
\newcommand{\rhokpr}{\rho_{\risk,\rx,\riselem}}
\newcommand{\gammatr}{\gamma_{\rx,\tx}}
\newcommand{\akpr}{\alpha_{\risk,\rx,\riselem}}
\newcommand{\bktr}{\beta_{\rx,\tx}}
\newcommand{\Gt}{G_{\tx}}
\newcommand{\Gr}{G_{\rx}}
\newcommand{\Gc}{G_c}

\newcommand{\tkp}{\risparamel_{\risk,\riselem}}

\newcommand{\rhorrk}{\rho_{\risk, \rr}}

\newcommand{\narrkp}{\bar{a}_{\risk, 2\rr+1}}

\newcommand{\nyrkpp}{{\bar{y}_{\risk, 2\rr+1}}}

\newcommand{\nnrkpp}{{\bar{n}_{\risk, 2\rr+1}}}
\newcommand{\nyrk}{{\bar{y}_{\risk, \rx}}}
\newcommand{\nyrkk}{{\bar{y}_{\risk, 2\rr}}}
\newcommand{\nark}{{\bar{a}_{\risk, \rx}}}
\newcommand{\narrk}{{\bar{a}_{\risk, 2\rr}}}
\newcommand{\nnrk}{{\bar{n}_{\risk, \rx}}}
\newcommand{\nntilderk}{\bar{n}_{\risk, 2\rr}}
\newcommand{\yrk}{{y}_{\risk, \rx}}
\newcommand{\yrkk}{{y}_{\risk, 2\rr}}
\newcommand{\ark}{{a}_{\risk, \rx}}

\newcommand{\nrk}{{n}_{\risk, \rx}}

\newcommand{\ssk}{\mathbf{s}_{\tempo}}

\newcommand{\bwk}{\mathbf{w}_{\tempo}}

\newcommand{\sskm}{\mathbf{s}_{\tempom}}
\newcommand{\sskp}{\mathbf{s}_{\tempop}}
\newcommand{\bzk}{\mathbf{o}_{\tempo}}
\newcommand{\bzkp}{\mathbf{o}_{\tempop}}
\newcommand{\betak}{\boldsymbol{\eta}_{\tempo}}
\newcommand{\Ns} {N_{\mathrm{s}}}

\newcommand{\Rk}{\mathbf{R}_{\tempo}}
\newcommand{\Rkp}{\mathbf{R}_{\tempop}}

\newcommand{\tra}{{T}}

\newcommand{\bsso}{\mathbf{s}_0}
\newcommand{\bmo}{\mathbf{m}_{0}}
\newcommand{\Po}{{\boldsymbol{\Sigma}_{0}}}
\newcommand{\vk}{\mathbf{v}_{\tempo}}

\newcommand{\bmkkmo}{\mathbf{m}_{\tempo \lvert \tempom}}
\newcommand{\bmkkp}{\mathbf{m}_{\tempop \lvert \tempo} }
\newcommand{\bmkkpp}{\mathbf{m}_{\tempop \lvert \tempop} }
\newcommand{\bhkkmo}{\mathbf{h}_{\tempo \lvert \tempom}}
\newcommand{\bhkkp}{\mathbf{h}_{\tempop \lvert \tempo}}
\newcommand{\bSkkmo}{\mathbf{S}_{\tempo \lvert \tempom}}
\newcommand{\Kk}{\mathbf{K}_{\tempo}}
\newcommand{\Kkp}{\mathbf{K}_{\tempop}}

\newcommand{\Skp}{\mathbf{S}_{\tempop}}
\newcommand{\Pkk}{\boldsymbol{\Sigma}_{\tempo\rvert \tempo}}
\newcommand{\Pkkmo}{\boldsymbol{\Sigma}_{\tempo \lvert \tempom}}

\newcommand{\Pkpok}{\boldsymbol{\Sigma}_{\tempop \lvert \tempo}}

\newcommand{\Hk}{\mathbf{J}_{\tempo}}
\newcommand{\Hkp}{\mathbf{J}_{\tempop}}

\newcommand{\bmkk}{\mathbf{m}_{\tempo \lvert \tempo}}
\newcommand{\Q}{\mathbf{P}}
\newcommand{\A}{\mathbf{T}}
\newcommand{\hssk}{\hat{\mathbf{s}}_{\tempo}}
\newcommand{\hsskp}{\hat{\mathbf{s}}_{\tempop}}

\newcommand{\Reakrr}{\Re\left\{\bar{a}_{\risk, 2\rr}\right\}}
\newcommand{\Reakrrp}{\Re\left\{\bar{a}_{\risk, 2\rr+1}\right\}}
\newcommand{\Imakrr}{\Im\left\{\bar{a}_{\risk, 2\rr}\right\}}
\newcommand{\Imakrrp}{\Im\left\{\bar{a}_{\risk, 2\rr+1}\right\}}

\newcommand{\txparamk}{\boldsymbol{\vartheta}_{\risk}}
\newcommand{\txparam}{\boldsymbol{\vartheta}}
\newcommand{\risparamel}{c}
\newcommand{\RISPARAM}{\mathbf{\MakeUppercase{\risparamel}}}
\newcommand{\risparam}{\mathbf{\risparamel}}
\newcommand{\RISPARAMk}{\RISPARAM_{\risk}}
\newcommand{\risparamk}{\risparam_{\risk}}
\newcommand{\precparamk}{\mathbf{f}_{\risk}}

\newcommand{\powerk}{P_{\risk}}
\newcommand{\innkp}{\mathbf{v}_{\tempop}}
\newcommand{\betakp}{\boldsymbol{\eta}_{\tempop}}

\newcommand{\powertx}{P_{\text{tx}}}

\begin{document}
 \bstctlcite{IEEEexample:BSTcontrol}

\title{{Two-Timescale Joint Precoding Design and \\ RIS Optimization for User Tracking in \\ Near-Field MIMO Systems}}

\author{Silvia Palmucci,~\IEEEmembership{Student Member,~IEEE,} Anna Guerra,~\IEEEmembership{Member,~IEEE,} \\ Andrea Abrardo,~\IEEEmembership{Senior Member,~IEEE,}  and Davide Dardari,~\IEEEmembership{Senior Member,~IEEE.} \\
\thanks{This work was sponsored, in part, by Theory Lab, Central Research
Institute, 2012 Labs, Huawei Technologies Co.,Ltd.\\ {This work was also sponsored by the European Union under the Italian National Recovery and Resilience Plan (NRRP) of NextGenerationEU, partnership on “Telecommunications of the Future” (PE00000001 - program “RESTART”).} \\S. Palmucci and A. Abrardo are with the University of Siena, Italy. E-mail: silvia.palmucci@student.unisi.it, abrardo@diism.unisi.it. A. Guerra is with the National Research Council of Italy, IEIIT-CNR, Italy. E-mail: anna.guerra@cnr.it. D. Dardari (corresponding author) is with the WiLAB - Department of Electrical and Information Engineering ``Guglielmo Marconi" - CNIT, University of Bologna, Italy. E-mail: davide.dardari@unibo.it. }
}

\markboth{Submitted to IEEE Transactions on Signal Processing}%
{S. Palmucci \MakeLowercase{\textit{et al.}}}
\maketitle
\begin{abstract}

In this paper, we propose a novel framework
that aims to jointly design the reflection coefficients of multiple \acp{RIS} and the precoding strategy of a single \ac{BS} to optimize the {self}-tracking of the position and the velocity of a single multi-antenna \ac{UE} {that moves either in the far- or near-field {region}.} Differently from the literature, and to keep the overall complexity
affordable, we assume that \ac{RIS} optimization is performed less frequently than localization and precoding adaptation. {{The proposed} procedure leads to minimize the inverse of the received power in the \ac{UE} position uncertainty {area,} between two subsequent optimization steps.
}
The optimal \ac{RIS} and precoder strategy is compared with the classical beam focusing strategy and {with a scheme} that maximizes the communication rate. It is shown that if the \acp{RIS} are optimized for communications, {their configuration} 
is suboptimal when used for tracking purposes.
{Numerical results show that in typical indoor environments with only one {BS} and a few {RISs} operating on millimeter waves, high location accuracy in the range of less than half a meter can be achieved.}

\end{abstract}

\acresetall
\section{Introduction}\label{Intro}


\IEEEPARstart{L}OCATION AWARENESS is an essential feature of sixth-generation cellular communication systems as high-accuracy positioning facilitates low latency and reliable wireless communications, also in indoor or GNSS-deprived environments \cite{de2021convergent}. Indeed, in harsh \ac{EM} environments achieving 
high positioning accuracy is a challenging task due to the presence of multipath, cluttering, and \ac{NLOS}  propagation \cite{wang2022location}. Traditional solutions to this issue include the deployment of numerous \acp{BS} and the use of advanced signal processing techniques \cite{witrisal2016high} while lower complexity alternatives include passive relays typically made of a single omnidirectional antenna.
More recently, the introduction of intelligent surfaces (ISs), either in active or reflective mode, has been welcomed as a game-changer low-power technology to realize \acp{SRE}\cite{he2022beyond}. Thanks to their capability to control the amplitude and phase of the impinging wavefront, \acp{RIS} can be used to boost communication performance by focusing the power on interested \acp{UE} and by enabling higher data rates \cite{renzo2019smart, basar2019wireless, alexandropoulos2021reconfigurable, flamini2022towards, zhang2022beam}. At the same time,  thanks to the possibility of constructing such surfaces with large physical apertures that enable an increased angular resolution, they also improve the accuracy of localization systems  \cite{wymeersch2020radio, elzanaty2021reconfigurable, zhang2021metalocalization, dardari2021nlos,guidi2021radio, wymeersch2022radio, chen2022tutorial}.  

The potential benefits brought by the adoption of \acp{RIS} in localization systems have been investigated in several papers. For example, in \cite{bjornson2022reconfigurable}, a \ac{SISO} \acf{OFDM} downlink scenario with a \ac{RIS} in a far-field regime is analyzed from communication and localization perspective. Similarly, the authors in \cite{wymeersch2020radio} derive \ac{CRLB} on positioning in a downlink scenario, demonstrating the benefits of using \acp{RIS} in far-field over methods that merely utilize the environment's natural scattering. 
Instead, the authors in \cite{elzanaty2021reconfigurable} explore the localization performance limits for an uplink \ac{MIMO} system operating using near-field propagation models, valid for limited distances, demonstrating the improvement in localization accuracy when a single \ac{RIS} is present. In \cite{guidi2021radio}, the impact of an \ac{EM} lens on near-field positioning performance is investigated through a maximum likelihood estimation analysis, whereas, in \cite{dardari2021nlos}, the problem of single-anchor localization assisted by \ac{RIS} is considered for applications characterized by frequent \ac{NLOS} conditions. Other practical \ac{RIS}-aided localization algorithms can be found in \cite{zhang2021metalocalization,zhang2020towards,nguyen2021wireless}. 

The localization and communication performance severely depends on how \acp{RIS} are designed and controlled. In this regard, a rich literature is available for optimizing \ac{RIS}-aided communications under different \ac{CSI} and/or {prior} localization knowledge \cite{abrardo2021intelligent,zhou2022joint,dai2021statistical}. \ac{RIS}  optimization schemes designed to improve communications are not, in general, optimum or good to increase localization accuracy because the respective performance indicators (e.g., the achievable rate vs. position error bound) depend differently on RIS configuration, channel characteristics, and geometry. Fewer results on \ac{RIS} optimization dedicated to localization are available.  
In particular, \cite{denis2021method} develops a method for the computation of the \ac{UE} position through multiple \acp{RIS} that are alternatively enabled and tuned using a localization-based cost function. The overall problem is expressed as an iterative procedure during which a minimum distance criterion is used to activate a specific \ac{RIS} and to estimate the \ac{UE} position. In \cite{rahal2022constrained}, the authors present a \ac{RIS} design method based on optimizing a closed-form \ac{PEB} in the presence of \ac{NLOS} and considering practical hardware constraints. In \cite{gao2022wireless}, a worst-case localization design is proposed based on the minimization of the squared \ac{PEB}. 
In \cite{feng2021power}, an optimization problem for \ac{RIS} localization and transmit power minimization is presented both for the case of single and multiple targets, using the \ac{CRLB} and the semidefinite release method for the power optimization problem.

Recently, the localization and communication tasks have been addressed jointly. 
Along this direction, the paper \cite{FanJiang2022} considers the localization statistics, determined using the \ac{FIM} of the observation model, and proposes a new framework for integrated localization and communication. Therefore, the subsequent \ac{RIS} configurations are fixed along a location coherence interval, while the \ac{BS} precoders are optimized at each channel coherence interval. In \cite{palmucci2022}, a joint communication and localization approach is presented in which the \ac{RIS} is designed to maximize the average rate by taking advantage of a \ac{RIS}-aided tracking procedure. The problem of jointly designing localization and communication is also tackled in \cite{he2020adaptive}, where a hierarchical codebook for the \ac{RIS} phase profile is proposed to enable adaptive bisection search over the angular space.

Most of the works entailing \ac{RIS}-aided localization consider snapshot positioning in the far-field region \cite{wang2022location, denis2021method, elzanaty2021reconfigurable, dardari2021nlos, feng2021power,zhang2021metalocalization,he2020adaptive}. In contrast, only a few works consider the \ac{UE} tracking problem. 
Among them, in \cite{boyu2022bayesian}, a multiple-antenna \ac{BS} estimates the position of a multiple-antenna \ac{UE} and tracks its trajectory through a message-passing algorithm that infers the \ac{UE} position starting from the estimated \acp{AOA}.
Such an observation model is valid when the system operates in the far-field and only bearing information can be retrieved to infer the \ac{UE} position. Moreover, two-step positioning approaches starting from intermediate parameters, e.g., \acp{AOA}, are suboptimal with respect to direct algorithms and require a proper characterization of the measurement statistics. In the algorithm proposed in \cite{boyu2022bayesian}, both the \ac{BS} and the \ac{RIS} design are optimized to minimize the Bayesian \ac{CRLB}. However, the \ac{RIS} optimization, designed to focus the reflected power towards the position estimate, does not consider the \ac{UE} uncertainty area. Moreover, the authors assume that, at each time step and whenever a new position estimate is available, the transmission parameters are optimized, leading to higher system latency and signaling overheads.
In general, better localization performance can be accomplished through directional beamforming at the \ac{BS} side, provided it is equipped with many antennas and, for the optimal result, that there exists a prior \ac{UE} location information \cite{fascista2022ris}. Hence, to increase the system degrees of freedom in a \ac{RIS}-aided environment, the optimization process may involve the joint design of \ac{BS} precoder and \ac{RIS} profiles, as it happens in \cite{boyu2022bayesian}. However, considering that most applications require a high localization update during the tracking process (e.g., $10$-$100\,\mathrm{Hz}$ refresh rate), it is obvious that the optimization of the \ac{RIS} at the same rates of location update and precoder optimization {could be demanding. Indeed, from the literature, it is well known that performing frequent \ac{RIS} optimization through channel estimation }would pose severe issues in terms of signaling overhead, complexity, latency, and technology {and different works have tackled the problem through a two-time scale approach, e.g., \cite{han2019large, hu2021two, zhi2022two}}. Therefore, { 
{in this paper},} 
{we will consider} 
 { two-timescale} solutions ensuring reliable tracking with much lower \ac{RIS} reconfiguration rates in the presence of \ac{NLOS} conditions between the \ac{BS} and the \ac{UE}, exploiting the opportunities of near-field propagation.  

\paragraph{Notation} Scalar variables, vectors, and matrices are represented with lower letters, lower bold letters, and capital bold letters, respectively (e.g., $x$, $\mathbf{x}$, and $\mathbf{X}$). The symbols $\left( \cdot \right)^T$, $\left( \cdot \right)^H$, and $\left( \cdot \right)^{-1}$ represent the transpose, conjugate transpose, and inverse operators of their arguments, respectively. 
We use $\tempo$ for discrete temporal indexing, $\tx$, $\rx$ for \ac{BS} and \ac{UE} antenna indexing, $\risk$ is the index for indicating a generic \ac{RIS} and $\riselem$ refers to a single radiating element of a \ac{RIS}. 
The notation $\mathbf{x}_{a \lvert b}$ indicates the value of a vector $\mathbf{x}$ at time instant $a$ estimated by considering the measurements collected up to time instant $b$. For example, $\mathbf{x}_{\tempo \lvert \tempom}$ is the value of $\mathbf{x}$ predicted at time instant $\tempom$ for the next time instant $\tempo$, whereas, once a new measurement becomes available at $\tempo$, this value is updated to $\mathbf{x}_{\tempo \lvert \tempo}$.  
{We denote with $f_{\mathbf{x}}(\boldsymbol{x})$ the probability density function of the random vector $\mathbf{x}$, and with $\mathbb{E}_{\mathbf{x}}$ its expectation.}
The symbol $\Theta = \left(\theta,\phi\right)$ includes both the elevation and azimuth {angles}. 

\section{{Main Contributions}}
\label{Main_Contributions}
{
To overcome the limitations present in the literature, in this work, for the first time, we study a two-time scale joint optimization problem for the \ac{BS} precoder and the \acp{RIS} reflection coefficients in a single-\ac{UE} \ac{MIMO} scenario where a moving \ac{UE}, either in the far- or near-field, estimates its trajectory using multiple \acp{RIS}.  The main novelty lies in the \ac{RIS} optimization procedure that is performed  to minimize  the tracking error in an uncertainty area that accounts for position uncertainty between two subsequent optimization steps. As a result, rather than maximizing the energy towards the instantaneous \ac{UE} position, as it is usually done, the \ac{RIS} optimization process leads to  minimizing the inverse of the received power in the \ac{UE} uncertainty area.  Moreover,  differently from the literature, the overall complexity is kept affordable, assuming that \ac{RIS} optimization is performed less frequently than localization and precoding adaptation. Nonetheless, this optimization accounts for the trajectory prediction within a finite time horizon up to the next optimization step, avoiding the so-called {\em deafness problem} caused by a loss of tracking due to the narrow beams that would result using a conventional RIS optimization.
\\
Toward this aim, the main proposed strategies and the corresponding results are summarized in the following.
\begin{itemize}
\item A novel observation model is introduced for the tracking process by considering a near-field narrowband signal model and direct positioning. According to this model, the observations are given by the product of the normalized received signals at adjacent \ac{UE} receiving antennas. In this way, they carry both bearing and ranging information in the near-field region \cite{guerra2021near}. Such an observation model allows strong non-linearities caused by phase periodicity to be mitigated, and the synchronization issues are avoided without affecting the localization accuracy. Hence, a classical low-complexity \ac{EKF} algorithm can be adopted \cite{dardari2015indoor}.
\item  By relying on \ac{BD} beamforming, we propose a pilot-based precoding method aimed at transmitting separated weighted beams toward the RISs, capable of removing both the cross-RIS interference at the \ac{UE} and the contribution of the direct BS-UE \ac{NLOS} connection, notoriously detrimental for localization.  
\item 
In order to be able to account for the area in which the UE is supposed to move between two subsequent RIS optimization steps, the \ac{RIS} optimization uncertainty area is obtained through a Gaussian Mixture Model-like approach, derived according to the motion model and the \ac{UE} estimated position at the \acp{RIS} optimization instants. 
\item Due to the high complexity and non-convexity of the derived function, the joint optimization problem for the transmission parameters based on the definition of the \ac{MSE} of the position estimate is recast in a simple form that allows optimization of the \acp{RIS} phase profiles and the precoders separately through an iterative \ac{BCD} algorithm that provides a local optimum.
\item The \ac{BS} precoders aims at finding the optimal weights for the transmitting beams. Hence, by following the same \ac{MSE} minimization approach as for \ac{RIS} optimization, we found that the optimal power allocation for tracking allows for a fair distribution of the power among the different \acp{RIS} so that the spatial diversity, beneficial for localization, is preserved. This differs from the idea of water-filling, used for maximizing the achievable communication rate, where most of the power is distributed in favor of the channels with the highest SNR, in a less fair manner.
\item In the numerical results, it is shown that in a typical millimeter-wave indoor scenario, a \ac{UE} can be tracked with high accuracy with a single \ac{BS}  and few \acp{RIS}, differently from other existing works that entail the use of many \acp{BS} to achieve the 5G positioning requirements. Moreover, it is confirmed that focusing, which is the typical strategy for \ac{RIS} configuration in the near-field regime,  is not the optimal strategy for tracking and that the  \ac{RIS} and precoder optimization process is different from that studied in the literature for communication that might result to be suboptimal when dealing with \ac{UE} tracking.
\end{itemize}
}

\section{Problem Formulation and System Geometry}
\label{sec:problem}

\subsection{Localization Scenario and Geometry}

We consider a localization and communication scenario in which a single \ac{UE} in position $\prx \in \mathbb{R}^3$ moves in the environment and localizes itself by processing the received signals sent by a single \ac{BS} in $\ptx \in \mathbb{R}^3$ and reflected by multiple \acp{RIS}. Either the \ac{UE} and the \ac{BS} are equipped with multiple antennas. We denote by $\prxr \in \mathbb{R}^3$ and $\ptxt \in \mathbb{R}^3$, respectively, the positions of the $r$th antenna on the \ac{UE} side, with $r \in \mathcal{R} \triangleq \left\{0, 1, \ldots, \Nrx-1\right\}$, and the $\tx$th antenna on the \ac{BS} side, with $\tx \in \mathcal{T} \triangleq \left\{0, 1, \ldots, \Ntx-1\right\}$. In the following, for simplicity, we denote $\mathbf{p}_{\text{RX},0}=\pRX$ and $\mathbf{p}_{\text{TX},0}=\ptx$. The considered scenario is also characterized by $K$ large \acp{RIS}, each of them comprising $P$ unit cells. Consequently each \ac{RIS} cell position is indicated as $\pkp \in \mathbb{R}^3$, $k \in \mathcal{K} \triangleq \left\{1, 2, \ldots, K \right\}$ and $p \in \mathcal{P} \triangleq \left\{0, 1, \ldots, P-1 \right\}$. 
\begin{figure}[t!]
\centering
\psfrag{b}[lc][lc][0.5]{$\beta$}
\psfrag{g}[lc][lc][0.5]{$\gamma$}
\psfrag{a}[lc][lc][0.5]{$\alpha$}
\psfrag{B}[lc][lc][0.5]{BS}
\psfrag{gg}[lc][lc][0.9]{}
\psfrag{c}[rc][rc][0.9]{} 
\psfrag{cc}[rc][rc][0.9]{}
\psfrag{h}[lc][lc][0.9]{} 
\psfrag{hh}[lc][lc][0.9]{} 
\psfrag{i}[rc][rc][0.9]{}
\psfrag{n}[lc][lc][0.9]{} 
\psfrag{nn}[lc][lc][0.9]{} 
\psfrag{d}[c][c][0.9]{$\tempo$}
\psfrag{e}[c][c][0.9]{$\tempop$}
\psfrag{f}[c][c][0.9]{$\mathsf{\tempo+2}$}
\psfrag{GG}[c][c][0.5]{\textit{Array Geometry}}
\psfrag{RIS}[lc][lc][0.5]{$k$th {RIS}}
\psfrag{MS}[lc][lc][0.5]{{UE}}
\psfrag{x1}[rc][rc][0.5]{$X$}
\psfrag{y1}[rc][rc][0.5]{$Y$}
\psfrag{z1}[rc][rc][0.5]{$Z$}
\psfrag{x}[rc][rc][0.5]{$X$}
\psfrag{y}[rc][rc][0.5]{$Y$}
\psfrag{z}[rc][rc][0.5]{$Z$}
\psfrag{bo}[rc][rc][0.5]{$\mathbf{p}_{\text{TX},0}$}
\psfrag{pb}[lc][lc][0.5]{$\pt$}
\psfrag{m}[lc][lc][0.5]{$\pr$}
\psfrag{mo}[lc][lc][0.5]{$\mathbf{p}_{\text{RX},0}=\mathbf{p}$}
\psfrag{lo}[rc][rc][0.5]{$\mathbf{p}_{k,0}$}
\psfrag{lk}[c][c][0.5]{$\mathbf{p}_{k,p}$}
\psfrag{dBL}[lc][lc][0.5]{$\dTXk$}
\psfrag{dBM}[lc][lc][0.5]{$d_{\text{RX}, \text{TX}}$}
\psfrag{dLM}[lc][lc][0.5]{$\dRXk$}
\psfrag{dlb}[rc][rc][0.5]{\textcolor{blue}{$d_{\risk,\riselem, \tx}$}}
\psfrag{dbm}[rc][rc][0.5]{\textcolor{blue}{$d_{\rx, \tx}$}}
\psfrag{dlm}[rc][rc][0.5]{\textcolor{blue}{$d_{\risk,\rx, \riselem}$}}
\psfrag{tb}[c][c][0.5]{$\theta_{\text{S},s}$}
\psfrag{p}[c][c][0.5]{$\phi_{\text{S},s}$}
\psfrag{do}[c][c][0.5]{$\mathbf{p}_{\text{S},0}$}
\psfrag{d}[c][c][0.5]{$\mathbf{p}_{\text{S},s}$}
\psfrag{db}[c][c][0.5]{$\mathsf{d}_{\text{S},s}$}
\includegraphics[width=\columnwidth,draft=false]
{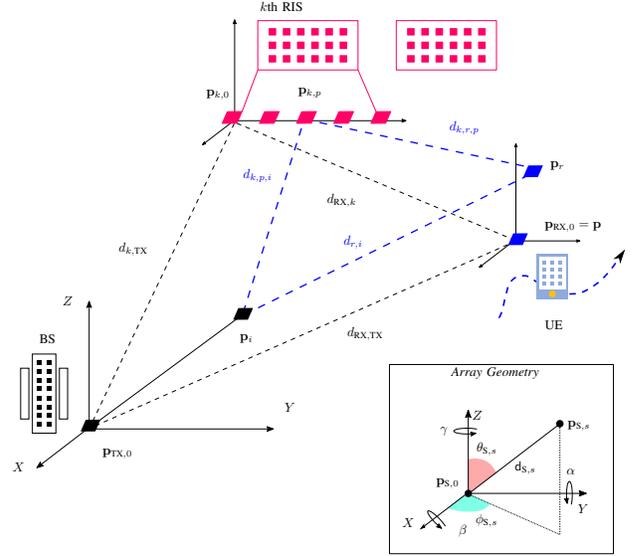}
\caption{Joint user tracking and communication scenario. }
\label{fig:scenario}
\end{figure} 
Let define the direction vector as
\begin{align}
 \mathbf{a}\left(\theta, \phi\right)& =\left[ \sin(\theta) \,\cos(\phi),\, \sin(\theta) \,\sin(\phi),\, \cos(\theta) \right]^T,
\end{align} where $\theta$ and $\phi$ are the elevation (measured from the $z$-axis to the $(x,y)-$plane) and azimuth (measured from the $x$-axis) angles. Then, considering the \ac{BS} as the center of the coordinate system, for each $\text{S} \in \left\{\text{TX}, \text{RX}, \risk \right\}$ and for each corresponding antenna index $s \in \left\{\tx, \rx, \riselem \right\}$, we can indicate the antenna coordinates of each array as \begin{align}\label{eq:antennacoord}
 &\mathbf{p}_{\text{S},s} = [x_{\text{S},s},\, y_{\text{S},s},\, z_{\text{S},s} ]^T= 
 {\ptx} + \mathsf{d}_{\text{S},s} \, \mathbf{a}\left(\theta_{\text{S},s}, \phi_{\text{S},s}\right), 
\end{align} where $\mathsf{d}_{\text{S},s}~=~\lVert \mathbf{p}_{\text{S},s} -\ptx \rVert_2$ is the distance from the \ac{BS}, and $\phi_{\text{S},s}~=~\operatorname{atan}\left((y_{\text{S},s} -{y_{\text{TX}}})/(x_{\text{S},s} -{x_{\text{TX}}}) \right)$,\footnote{The operator $\operatorname{atan}$ corresponds to the four-quadrant inverse tangent.} and $\theta_{\text{S},s}~=~\operatorname{acos}\left((z_{\text{S},s} -{z_{\text{TX}}})/ \mathsf{d}_{\text{S},s} \right)$
are the azimuth and elevation angles.\footnote{{In the next, the subscripts RX and TX are omitted and $\left\{\pr, \dr, \thetar, \phir\right\}$ and $\left\{\pt, \dt, \thetat, \phit\right\}$ are used instead.}} Starting from \eqref{eq:antennacoord}, the distance between the $\tx$th antenna of the \ac{BS} 
and the $p$th element of the $k$th \ac{RIS} is given by\footnote{In the sequel, we indicate $\dTXRX \triangleq \left \| \prx - \ptx \right \| $, $\dtr \triangleq \left \| \prxr - \ptxt \right \| $, $\dTXk \triangleq \left \| \pk - \ptx \right \| $, and $\dRXk \triangleq \left \| \prx - \pk \right \| $.}
    \begin{align}
    \label{eq:inter_d}
        &\dktp= \lVert \pt - \pkp \rVert_2 =  \left\{\dt^2 + \dkp^2-2\, \dt\, \dkp\, \left[ \sin(\thetat)\,\right.\right.  \cdot \nonumber \\
         & \left. \left. \qquad \times \sin(\thetakp)\,\cos(\phit- \phikp) + \cos(\thetat)\, \cos(\thetakp) \right] \right\}^{1/2}
    \end{align}
where 
$\left\{\dt, \thetat, \phit\right\}$ are the spherical coordinates of the generic 
$\tx$th 
transmitting antenna, as defined above.  Analogously, we indicate the angle between the $\tx$th antenna of the \ac{BS} 
and the $p$th element of the $k$th \ac{RIS} as \cite{abrardo2021intelligent}
$\Thetaktp=\left (\thetaktp,\phiktp \right )$.
$\dkpr$ and $\Thetakpr$ are computed in a similar manner for every $k$th \ac{RIS}-\ac{UE} antenna couple. 
Next, we consider that the \ac{BS} and \ac{RIS} antenna coordinates and array orientation are known. 
By contrast, the position of the \ac{UE} is unknown, whereas its orientation is considered known (e.g., estimated by onboard sensors such as compass and gyroscopes {\cite{dardari2011satellite, guidi2015personal}}). {In the numerical results, we will show the impact of orientation residual errors in the performance.} An example of the used localization and communication scenario and its geometry is reported in Fig.~\ref{fig:scenario}. From \cite{abrardo2021intelligent, dardari2021nlos} the local reflection coefficient of a unit cell of \ac{RIS} can be written as ${r}_{\risk, \rx, \riselem, \tx}  = G_\mathrm{c} e^{j \Psi_{\risk,\riselem}} \sqrt{F\left( \Theta_{\risk, \riselem, \tx} \right) F\left(\Thetakpr \right)},$
where $G_c$ is the gain of the unit-cell and $e^{j \Psi_{\risk,\riselem}}$ is the unit-norm load reflection coefficient. $F ( {\Theta} )$ denotes the normalized power radiation pattern of each unit cell, which is frequency-independent within the bandwidth of interest and modeled as an exponential-Lambertian radiation pattern with parameter $q$ as
\begin{align}
F\left( {\Theta} \right) = \left\{ 
\begin{array}{ll}
\cos^q \left( \theta \right), & \theta \in \left[0, \pi/2 \right], \phi \in \left[0, 2\pi \right] \\
0, & \text{otherwise}
\end{array}.
\right. \label{eqnprp}
\end{align} 

\subsection{Pilot Signal Model}
In the following, we describe the proposed pilot-based method for localization and derive the corresponding signal models. The proposed method aims at isolating the contribution of the different \acp{RIS} at the \ac{UE} receiver and eliminating the contribution of the direct BS-\ac{UE} channel since, in many cases, it is a \ac{NLOS} contribution that does not contain useful information for localization. The \ac{BS} pilot signal is
\begin{equation}
 \mathbf{X}= \mathbf{F}\, \mathbf{Q} \in \mathbb{C}^{{\Ntx} \times L},
 \label{eq:pilot}
\end{equation}
where $\mathbf{Q}=[\mathbf{q}_1, \ldots, \mathbf{q}_K ]^H \in \mathbb{C}^{K \times L}$ contains $K$ sequences of unitary envelope orthogonal pilot signals, i.e. $|{q}_{k,l}|~=~1$, $\mathbf{q}_k^H \mathbf{q}_m=L\,\delta_{k,m}$, with $\mathbf{q}_k \in \mathbb{C}^{L \times 1}$, $L \ge K$ being the length of the pilot sequence.\footnote{More precisely, $L = K$ is the minimum pilot length to obtain orthogonality, although a higher $L$ can also be considered to reduce the overall noise of the received signal at the expense of communication overhead.} The precoding matrix is  indicated as $ \mathbf{F}=[\mathbf{f}_1, \ldots, \mathbf{f}_K ] \in \mathbb{C}^{\Ntx \times K}$ and each columns $\precparamk=\left\{ \ftk \right\} \in \mathbb{C}^{\Ntx \times 1}, \, k \in \mathcal{K},$ is the precoding vector for the $k$th \ac{RIS}. The overall received signal at the \ac{UE}, $\mathbf{Y}\in \mathbb{C}^{\Nrx \times L}$,  is
\cite{abrardo2021intelligent, elzanaty2021reconfigurable} 
\begin{align}
 \mathbf{Y}&= \mathbf{H} \mathbf{X} + \sum_{k=1}^{K} \mathbf{B}_k \RISPARAMk \mathbf{G}_k \mathbf{X} +\mathbf{N}\triangleq \mathbf{H} \mathbf{X} + \sum_{k=1}^{K} \mathbf{Z}_{k}+\mathbf{N},
 \end{align}
 where $\mathbf{H}= \left\{ \htr\right\}\in \mathbb{C}^{\Nrx \times \Ntx}$ is the direct \ac{BS}-\ac{UE} channel matrix, $\mathbf{B}_k=\left\{ \bkpr \right\} \in \mathbb{C}^{\Nrx \times P}$ is the channel matrix between the $k$th \ac{RIS} and the \ac{UE}, $\mathbf{G}_k=\left\{\gktp \right\} \in \mathbb{C}^{P \times \Ntx}$ is the channel matrix between the \ac{BS} and the $k$th RIS, $\RISPARAMk=\operatorname{diag}\left(\risparamel_{\risk ,1}, \ldots, \tkp, \ldots, \risparamel_{\risk ,P}\right) = \operatorname{diag} \left(\risparamk\right) \in \mathbb{C}^{P \times P}$ contains the \ac{RIS}s responses, and $\mathbf{N} \in \mathbb{C}^{\Nrx \times L}$ represents the additive white Gaussian noise sequences with variance $\sigma^2$. For the communication system it is assumed $\Ntx \ge \Nrx$. The generic \ac{RIS} coefficient $\left\{\tkp\right\} \, , p \in \mathcal{P}, \, k \in \mathcal{K}$, denotes the normalized reflection coefficient of the \textit{p}th element of the \textit{k}th \ac{RIS}, i.e., $\tkp=e^{\jmath\, \Psi_{ \risk, \riselem}}$. Hence, the channel components are modeled as \begin{align}\label{eq:directch} 
 &\htr = \sqrt{\frac{\kappa_{h}}{\kappa_{h}+1}} \gammatr \,e^{ \left( - \jmath \frac{2\pi}{\lambda} \dtr \right )} + \sqrt{\frac{1}{\kappa_{h}+1}} \bktr, \\ 
 &\gktp = \rhoktp \,e^{ \left( - \jmath \frac{2\pi}{\lambda}\, \dktp \right )}, \\
& \bkpr = \sqrt{\frac{\kappa_b}{\kappa_b+1}} \rhokpr\,e^{ \left( - \jmath \frac{2\pi}{\lambda}\, \dkpr \right )} + \sqrt{\frac{1}{\kappa_b+1}} \akpr, \label{eq:cascadedch2}
\end{align} where $\gammatr \triangleq \frac{\lambda}{4\pi} \, \frac{\sqrt{\Gt \Gr}\, }{\dTXRX }$, $\rhoktp \triangleq \frac{\lambda}{4\pi} \, \frac{\sqrt{\Gt\,\Gc\, F(\Thetaktp)}  }{\dTXk }$, $\rhokpr \triangleq \left(\frac{\lambda }{4\pi} \, \frac{\sqrt{\Gr\, \Gc\, F\left( \Thetakpr \right)}\,  }{\dRXk }\right)$ are the channel amplitude of the \ac{LOS}  components, {${\kappa_{h} \ge 0}$} and ${\kappa_b \ge 0}$ are the Ricean factors {for the direct BS-\ac{UE} and \ac{RIS}-\ac{UE} links, respectively}, $\akpr \sim \mathcal{CN}(0,\rhokpr^2)$ {and $\bktr \sim \mathcal{CN}(0,\gammatr^2)$} denote the random complex fading coefficients of the \ac{NLOS} component \cite{cunhua2022overview}. $\rhokpr$, $\rhoktp$, and $\gammatr$ represent the compact notation of the amplitude channel coefficients (except for the {Ricean} factor terms), $\lambda$ is the wavelength, $\Gt$ is the gain of a single antenna element from BS, $\Gr$ is the gain of a single antenna element from UE, and $G_c$ is the gain of a \ac{RIS} unit cell. \footnote{{According to the observation model entailed for this work (see Sec. \ref{sec:tracking}-B) we can directly neglect BS-UE synchronization errors since they will not influence the algorithm performance.}} {In accordance to \cite{hou2020reconfigurable}, we assume a \ac{LOS} channel between the \ac{BS} and the \ac{RIS} which is a realistic condition of most practical scenarios.} The received signal is then projected over each pilot sequence $\mathbf{q}_k$ to get the contribution of the $k$th \ac{RIS}, as \begin{align}
\mathbf{y}_k&= \frac{1}{L} \mathbf{Y}\mathbf{q}_k= 
 \sum_{l=1}^{K} \mathbf{B}_l \RISPARAM_l \mathbf{G}_l \mathbf{f}_k + \mathbf{H} \mathbf{f}_k + \frac{1}{L} \mathbf{N} \mathbf{q}_k \nonumber \\
 &= \mathbf{z}_{k} +\mathbf{n}_k \in \mathbb{C}^{\Nrx \times 1},
 \label{eq:y_o}
 \end{align} where $ \mathbf{z}_{k} \triangleq \mathbf{B}_k \RISPARAMk \mathbf{G}_k \mathbf{f}_k $ represents the useful part of the received signal with respect to the \textit{k}th \ac{RIS}, $\mathbf{n}_k = \sum_{l=1, l\not = k}^{K}\mathbf{z}_{l}+\mathbf{H} \mathbf{f}_k
+ \mathbf{n}'_k$ is the noise plus interference terms, and $\mathbf{n}'_k \triangleq \frac{1}{L} \mathbf{N} \mathbf{q}_k$ is the AWGN noise term after correlation with the pilot sequence where each element has a power of $\sigma^2/L$, with $L$ being the pilot sequence length. 

\subsection{Block Diagonalization Precoding}\label{BDP}
The goal of the proposed precoding strategy is to maximize the received energy while eliminating interference to the $k$th received signal in \eqref{eq:y_o} due to the presence of the direct \ac{BS}-\ac{UE} channel and the other \acp{RIS}.\footnote{This goal can be achieved by placing the \ac{RIS} at a certain height above the ground and a sufficient distance from each other so that they receive energy only from the intended beam. Optimal \acp{RIS} placement is beyond the scope of this work and has already been studied in \cite{ghatak2021placement, wymeersch2020radio} for localization-based applications. In our settings, \ac{RIS} placement for simulations is chosen to ensure model assumptions.} In this case, 
$\sum_{l=1, l\not = k}^{K}\mathbf{z}_{l} \simeq 0$ and $\mathbf{H} \mathbf{f}_k \simeq 0$. 
 {
 Indeed, \ac{NLOS} conditions occur frequently in the direct \ac{BS}-\ac{UE} link in practical wireless communication scenarios. Moreover, this is the situation where the \acp{RIS} can provide the greatest benefits. Therefore, in our work, we assume that the \ac{LOS} component of the BS-UE signal is blocked, i.e., the BS-UE link only results in NLOS propagation, which is not useful for localization. Under these NLOS conditions, localization is indeed very difficult and subject to serious errors, \cite{dardari2009ranging, Sahinoglu2008ultra} and therefore the direct BS-UE connection is not considered in our localization scheme.}
Assuming a large number of antennas on the BS, this condition can be exactly satisfied by using \ac{BD} beamforming. This procedure, commonly used to suppress interference between different \acp{UE} \cite{Saggese2019}, aims to orthogonalize the available channels by projecting the transmitted signal for the $k$th \ac{RIS} into the null space of its interference. To elaborate, we introduce the \textit{k}th \ac{RIS} interference channel matrix $\tilde{\textbf{G}}_k \in \mathbb{C}^{\tilde{\alpha} \times \Ntx}$ as 
\begin{equation}
\tilde{\textbf{G}}_k=\left[\textbf{G}_1^T,\, \ldots, \, \textbf{G}_{{k}-1}^T,\, \textbf{G}_{{k}+1}^T, \, \ldots, \textbf{G}_K^T, \, \textbf{H}^T \right]^T ,
\end{equation} 
with $\tilde{\alpha}=\Nrx+(K-1)P$.
Applying the \ac{SVD} it is possible to rewrite the matrix as 
 \begin{equation}
    \tilde{\textbf{G}}_k= \tilde{\textbf{U}}_k\tilde{\boldsymbol{\Lambda}}_k\tilde{\textbf{V}}_k^H. 
\end{equation} 
{Denoting by} $\tilde{p} \triangleq \operatorname{rank}(\tilde{\boldsymbol{\Lambda}}_k)$, {we can write} $\tilde{\textbf{V}}_k= [ \tilde{\textbf{V}}_k^{(\mathrm{1})},\, \tilde{\textbf{V}}_k^{(\mathrm{0})}]$, {where $\tilde{\textbf{V}}_k^{(\mathrm{1})} \in \mathbb{C}^{\Ntx \times \tilde{p}}$ and $\tilde{\textbf{V}}_k^{(\mathrm{0})} \in \mathbb{C}^{\Ntx \times \Ntx-\tilde{p}}$ spans the null space of $\tilde{\textbf{G}}_k$.}
Specifically, the precoding vector $\mathbf{f}_k$ is designed as $\mathbf{f}_k = \tilde{\textbf{V}}_k^{(\mathrm{0})}\bar{\mathbf{f}_k}$ such that $\mathbf{G}_l \mathbf{f}_k = \mathbf{0}$, for $l \ne k$ and $\mathbf{H} \mathbf{f}_k = \mathbf{0}$, i.e., the inter-\ac{RIS} and direct link interferences are completely removed in \eqref{eq:y_o}. As for $\bar{\mathbf{f}}_k$, it represents the precoding vector of an equivalent MIMO channel with $\Ntx-\tilde{p}$ antennas, transmitting over the equivalent channel $\bar{\textbf{G}}_k=\textbf{G}_k \tilde{\textbf{V}}_k^{(\mathrm{0})} \in \mathbb{C}^{P \times (\Ntx-\tilde{p})}$, and can be designed by classical beamforming techniques. To elaborate, we denote by $\bar{\mathbf{v}}^{(\mathrm{0})}_k$ the first right singular vector of the SVD decomposition of $\bar{\textbf{G}}_k$, 
which corresponds to its optimal beamforming vector.
The normalized beamforming vector ${\mathbf{v}}^{(\mathrm{0})}_k$ for ${\textbf{G}}_k$ can be expressed as follows
\begin{align}\label{eq:norm_bm}
{\mathbf{v}}^{(\mathrm{0})}_k = \tilde{\textbf{V}}_k^{(\mathrm{0})} \bar{\mathbf{v}}^{(\mathrm{0})}_k / \lVert{\tilde{\textbf{V}}_k^{(\mathrm{0})} \bar{\mathbf{v}}^{(\mathrm{0})}_k}\rVert.
\end{align} 
Thus, if we denote by $\beta_k^2$ the power budget allocated at the \ac{BS} for the $k$th RIS, we have
\begin{equation}
   {{\mathbf{f}}_k ={\beta_k}{\mathbf{v}}^{(\mathrm{0})}_k}.
   \label{precoder}
\end{equation}
It is worth noting that the \ac{BD} method reduces the number of equivalent antennas available at the \ac{BS} for each \ac{RIS}, thus, limiting the beamforming gain. However, if the \acp{RIS} are at a certain height and sufficiently spaced, this effect may be negligible. Due to the elimination of the interference, the following expression for the received signal from the $k$th \ac{RIS} results:
\begin{align}\label{eq:rx_signal}
 y_{\risk, \rx} &= 
\sqrt{\frac{\kappa_b}{\kappa_b+1}} \sum_{\tx=0}^{\Ntx-1} \ftk\, \sum_{\riselem=0}^{P-1} \rhoktp \, \rhokpr \, \tkp \, e^{- \jmath \frac{2\pi}{\lambda}\, \dktpr} \nonumber \\
&+ n_{\risk, \rx} = a_{\risk, \rx}+n_{\risk, \rx},
\end{align} 
with $\dktpr \triangleq \dktp + \dkpr$. The noise $n_{\risk, \rx}$ includes the AWGN noise and the multipath components after correlation with the pilot signal and can be expressed as $n_{\risk, \rx} \sim \mathcal{CN}\left(0,{\sigma_{\risk, \rx}^2} \right) = \mathcal{CN}\left(0,\frac{\sigma^2}{L}+ \frac{1}{\kappa_b+1} \sum_{\riselem=0}^{P-1} \rhokpr^2 \right)$. Since in most practical situations, the dimension of the \ac{UE} antenna system is small compared to the distance to the RIS, we can assume that the same power is received at the different \ac{UE} antennas, i.e., $\rhokpr = \rho_{\risk , \riselem},\, \forall \rx \in \mathcal{R}$. Accordingly, we have the same noise variance at the different antennas, i.e., $n_{\risk,  r} \sim \mathcal{CN}\left(0,{\sigma_{\risk}^2} \right)$.

\section{Bayesian Localization: User Tracking}
\label{sec:tracking}
We now consider a single \ac{UE} tracking problem in \acp{SRE} where localization is aided by the presence of multiple \acp{RIS}. A common representation for tracking systems is a state-space model \cite{dardari2015indoor} consisting of a {\em transition model} that describes the evolution of the state over time, and an {\em observation model} that specifies how the measurements are related to the \ac{UE} state. In mathematical terms we have 
\begin{align} \label{eq:motionmodel}
&\ssk = f\left( \sskm\right) + \bwk, \\ 
& \bzk = h\left({\ssk} \right) + \betak, 
\label{eq:observationmodel}
\end{align} 
where $\ssk \triangleq \left[\pRXt, \, \dot{\mathbf{p}}_{\tempo} \right] \in \mathbb{R}^{\Ns}$ is the state vector at time instant $\tempo$ and contains the position ($\mathbf{p}_{\text{RX}}$) and velocity ($\dot{\mathbf{p}}_{\text{RX}}$) of the \ac{UE} at time $\tempo$, and $\Ns$ is the dimension of the state vector. The transition function is indicated as $f\left( \sskm\right)$ (in the following, a linear function is assumed as $f\left( \sskm\right)=\A\, \sskm$), and $\bzk$ is the observation model with $h\left({\ssk}\right)$ as the observation function. $\bwk$ and $\betak$ are stochastic Gaussian noise processes. 

\paragraph{Transition Model} For the transition model, we consider a second-order kinematic model \cite{bar2004estimation} 
\begin{align}\label{eq:tmodel}
&\A=\left[\begin{array}{cc} \mathbf{I}_3 & d\tempo\, \mathbf{I}_3 \\
\mathbf{0}_3 & \mathbf{I}_3
\end{array} \right],  \\
&\bwk\sim \mathcal{N} \left(\bwk; \mathbf{0}, \Q \right), && \Q=\left[\begin{array}{cc}
\frac{d\tempo^3}{3} \, \Q_{\mathsf{a}}& \frac{d\tempo^2}{2} \, \Q_{\mathsf{a}}\\
\frac{d\tempo^2}{2} \, \Q_{\mathsf{a}} & d\tempo \, \Q_{\mathsf{a}}
\end{array} \right], \nonumber
\end{align} 
where $d\tempo$ is the delay between two adjacent time instants $\tempo$ and $\tempop$, and $\Q_{\mathsf{a}} = \operatorname{diag}\left(\sigma^2_{\mathsf{a},\mathsf{x}}, \sigma^2_{\mathsf{a},\mathsf{y}} ,\sigma^2_{\mathsf{a},\mathsf{z}} \right)$ \cite{guerra2021near}. 

\paragraph{Observation Model}
\label{sec:obs_model}In MIMO systems, different signal parameters are considered to retrieve the location information \cite{elzanaty2021reconfigurable, dardari2015indoor}. In the presence of wideband signals, \ac{TOA} and \ac{AOA} can be used together {to allow positioning} \cite{guerra2018single}, while with narrowband signals and multiple antennas, \acp{AOA} {can be extrapolated from the signal phases in far-field conditions} \cite{bjornson2022reconfigurable}. {When operating in the near-field region, the phases of narrowband signals embodies richer information by also depending on ranging \cite{guerra2021near, guidi2021radio}.}

Such approaches, {that first derive the geometric signal parameters and then estimate the position starting from them}, are referred to as two-step techniques. Two-step approaches are generally suboptimal compared to direct methods that start from the received signal and avoid intermediate processing steps \cite{dardari2021nlos}. In addition, modeling the measurement noise in the two-step approach could be a difficult task due to the nonlinear operations, which is avoided when starting directly from the received signals whose statistics are known \cite{wymeersch2022radio, chen2022tutorial}. {Therefore,} in our narrowband method, we opt for a direct approach, where the range and angle information is implicitly accounted for by processing the received signals. 

Consequently, to define the observation vector, let denote the normalized received signals under the assumption of high \ac{SNR} as 
\begin{align}
 &\nyrk =\frac{\yrk}{|\yrk|} 
 \approx  \frac{\ark+\nrk}{\lvert a_{\risk} \rvert} = \nark+ \nnrk,
 \label{eq:rx_signal_norm}
\end{align} 
where in \eqref{eq:rx_signal_norm} we have exploited the assumption of equal power at the different \ac{UE} antennas, leading to $\lvert \ark\rvert = \lvert a_{\risk}\rvert$, and we have introduced the normalized useful term $\nark= \frac{\ark}{\lvert a_{\risk} \rvert}$. For the noise term we have $\nnrk \sim \mathcal{CN}\left(0,{\bar{\sigma}_{\risk}^2} \right)$
with 
\begin{equation}\label{eq:noisem}
\bar{\sigma}_{\risk}^2 \simeq \frac{\sigma_{\risk}^2}{\left|a_{\risk}\right|^2}.
\end{equation} 
Starting from the normalized received signals in \eqref{eq:rx_signal_norm}, we consider the correlation between measurements at adjacent antennas as observations. In particular, for $Z = \lfloor \Nrx/2 \rfloor$, the observation vector is given by 
\begin{align}\label{eq:obsRX_2}
\bzk \triangleq&\left[\Re\left\{\boldsymbol{\rho}_{1} \right\}, \ldots, \Re\left\{\boldsymbol{\rho}_k\right\}, \ldots, \Re\left\{\boldsymbol{\rho}_{K} \right\},  \nonumber \right.\\
&\left. \Im\left\{\boldsymbol{\rho}_{1} \right\}, \ldots, \Im\left\{\boldsymbol{\rho}_k\right\}, \ldots, \Im\left\{\boldsymbol{\rho}_{K} \right\} \right] \in \mathbb{R}^{2K\, Z}, 
\end{align} 
where the generic term  $\boldsymbol{\rho}_k=\left\{\rhorrk \right\}$, $\rr \in \mathcal{Z} \triangleq \left\{0,1, \ldots, Z-1\right\}$, is a phase gradient given by 
\begin{equation}\label{gradient}
\rhorrk= \left(\nyrkpp\right) \, \left(\nyrkk\right)^*. 
\end{equation} 
{When operating within the radiating near-field domain, \eqref{gradient} contains both range and bearing information (as it can also be seen from the system geometry of \eqref{eq:inter_d}), while in the far-field, it contains only \ac{AOA} information \cite{guerra2021near}.} Moreover, by directly processing the received signals, it is possible to fit the Gaussian model for the observation noise underlying the implementation of  \ac{KF} differently from what happens when considering the statistics of a specific estimator for \acp{AOA}. 

{Another advantage} of considering the gradient in \eqref{gradient} {is} the possibility of avoiding strong non-linearities arising from fast variations of the observed signals along the antenna array, which could undermine the functioning of the tracking algorithm considered in this paper. More precisely, \eqref{gradient} is used to achieve a double effect. On the one hand, normalizing the received signal (i.e., using $\nyrkk$ instead of $\yrkk$) allows the observations to be independent of the received powers, which contain information about the distance but can vary rapidly due to the ability of \acp{RIS} to form very narrow beams. Moreover, \ac{RSS} measurements lead to less accurate positioning with respect to \ac{TOA} or \ac{AOA} measurements \cite{elzanaty2021reconfigurable}. On the other hand, conjugate multiplication (i.e. $ \nyrkpp \, \nyrkk^*$ instead of $\nyrkpp$ alone) allows to get rid of the dependence on absolute phase, which is also rapidly variable {and affected by possible clock offsets deriving from synchronization errors. In practical terms the use of the phase differences instead of the absolute phases does not entail any loss of important information}.

The observation function in \eqref{eq:observationmodel} is given by the expected mean of \eqref{eq:obsRX_2}, i.e., $h\left({\ssk} \right) \triangleq \mathbb{E}\left[\bzk \right]$ whose generic elements can be easily computed as \begin{align}\label{eq:obs_fun_re_2}
&h\left( \Re\left\{\rhorrk \right\} \right)= \mathbb{E}\left\{\Re\left\{\rhorrk \right\} \right\}= 
\Re\left\{ \narrkp\, \narrk^* \right\} \\
&h\left( \Im\left\{\rhorrk \right\} \right)=\mathbb{E}\left\{\Im\left\{\rhorrk \right\} \right\}=  \Im\left\{ \narrkp \, \narrk^* \right\}. \label{eq:obs_fun_im}
\end{align} The measurement noise $\betak$ of \eqref{eq:observationmodel} is considered as zero-mean Gaussian distributed noise with a diagonal covariance matrix $\mathbf{R}_{\tempo}$. Now consider $\left \{ \mathbf{R}_{\tempo} \right \}_{n,n}$, the $n$th element of the main diagonal of $\Rk$, and denote by $\mathcal{I}_k = \left\{(k-1)Z+\rr+(m-1)KZ{+1}\right\}$, for $\rr \in \mathcal{Z}$, $m = 1,2$ the set of indices 
of the observation vectors corresponding to the signal received from the $k$th RIS.  Thus, considering a general $n \in \mathcal{I}_k$ and for each $\rr \in \mathcal{Z}$, we have
\begin{align}\label{eq:var_ok_2}
 \left\{\mathbf{R}_{\tempo}\right\}_{n,n} &= \text{var}\left\{\Re\left\{\rhorrk \right\} \right\}= \text{var}\left\{\Im\left\{\rhorrk \right\} \right\} \nonumber \\
 & \approx  \text{var} \left \{ \Re \left (\narrkp\right)\Re \left (\nntilderk\right) +\Im \left (\narrkp\right)\Im \left (\nntilderk\right)  \right. \nonumber \\
 &\left.+ \Re \left (\narrk\right) \Re \left (\nnrkpp\right)+ \Im \left (\narrk\right) \Im \left (\nnrkpp\right) \right \} \nonumber \\ 
 & = \frac{\bar{\sigma}_{\risk}^2}{2} \left[\lvert \bar{a}_{\risk}\rvert ^2\right] 
 +\frac{\bar{\sigma}_{\risk}^2}{2} \left[\lvert \bar{a}_{\risk}\rvert ^2\right]=\bar{\sigma}_{\risk}^2  = \frac{\sigma_{\risk}^2}{| a_{\risk} |^2} 
\end{align} 
where, as before, we assume a large \ac{SNR} regime and independence between the received signals at adjacent antennas. In words, $\mathbf{R}_{\tempo}$ has a blockwise diagonal structure in which all elements corresponding to the signal received from the $\risk$th \ac{RIS} have the same value $\bar{\sigma}_{\risk}^2$. 

\paragraph{EKF Algorithm}
Among the Bayesian estimators, we use the \ac{EKF}, which considers the signals in \eqref{eq:observationmodel}. 
The state is described by a Gaussian distribution $\ssk\sim \mathcal{N}\left(\ssk; \bmkk, \Pkk \right)$, where $\bmkk \, {\in \mathbb{R}^{\Ns}}$ and $\Pkk \, {\in \mathbb{R}^{\Ns^2}}$ are the posterior mean vector and the covariance matrix of the state. An important step of the \ac{EKF} is the evaluation of the Jacobian matrix associated with the linearization of the observation model $h \left({\ssk} \right)$, which can be written as $\Hk \triangleq \nabla_{\ssk}\, h \left({\ssk} \right)$, where $\nabla_{\ssk}$ is the gradient with respect to the state vector. The Jacobian is evaluated at $\ssk=\bmkkmo$, where $\bmkkmo$ is the predicted state (for $\tempo=1$ it is $\bmkkmo=\bmo$). The Jacobian computation is reported in the supplementary material as Appendix A, whereas the \ac{EKF} is reported in Alg. \ref{alg:EKF}, in which $\txparamk^{(\tempo)}$ represents the transmission parameters to be optimized at time $\tempo$ and for the \textit{k}th \ac{RIS}.  

\begin{algorithm}[h!]
\SetAlgoLined
{
\textbf{Initialization for $\tempo=0$}:\\
- Initialize the state $\bsso \sim \mathcal{N}\left(\bsso; \bmo, \Po \right)$; \\
- Initialize the transmission parameters
$\txparamk^{(1)}=(\risparamk^{(1)}, \beta^{(1)}_k)$ for $k = 1,\ldots,K$; \\
- Set $\mathbf{m}_{1 \lvert 0}=\bmo,\, \boldsymbol{\Sigma}_{1 \lvert 0}= \Po,\,\, \mathbf{h}_{1 \lvert 0}=h \left( \mathbf{m}_{1 \lvert 0} \right)$; \\
 \For{$\tempo=1, \ldots, \mathsf{T}$}{ 
 \textbf{Measurement update};\\
- Collect a new measurement vector $\bzk$; \\
- Compute the innovation: \\
 $\vk=\bzk-\bhkkmo$; \\ 
 $\bSkkmo=\Hk\Pkkmo\Hk^{\tra} + \Rk$; \\
- Compute the Kalman gain: \\
 $\Kk=\Pkkmo \Hk^{\tra}\,\bSkkmo^{-1}$; \\
- Update the posterior state estimate and covariance: \\
$\bmkk=\bmkkmo + \Kk\,\vk$;\\
$\Pkk=\Pkkmo-\Kk\,\bSkkmo\,\Kk^{\tra}$; \\
 \textbf{State Estimation};\\
- Estimate the state: \\ 
 $\hssk=\bmkk$; \\
 \textbf{Time Update: State Prediction and RIS Optimization}; \\
- Predict the a-priori moments of the state based on the previous a-posteriori estimates; \\ 
$\bmkkp = \A \, \bmkk$; \\
$\Pkpok = \A\, \Pkk \, \A^{\tra} + \Q$; \\
Evaluate $\mathbf{J}_{\tempop}$ and
$\bhkkp = h\left(\bmkkp\right)$; \\
- Calculate $\txparamk^{(\tempop)}=(\risparamk^{(\tempop)}, \beta^{(\tempop)}_k)$ for $k = 1,\ldots,K$ (see Alg. \ref{alg:RIS_Opt}-\ref{alg:RIS_and_Precoder_Opt}). \\
}
}
 \caption{Extended Kalman Filter}
 \label{alg:EKF}
\end{algorithm}

\subsubsection*{Noise covariance computation}
Calculating $\mathbf{R}_{\tempo}$ in \eqref{eq:var_ok_2} to be plugged into the EKF is problematic because ${\sigma}_{k}^2$ depends on the multipath, which is difficult to predict, and $| a_{\risk} |^2$ depends on the \ac{UE}'s location, which is the unknown term that must be estimated. To circumvent this difficulty, we consider an approximation for $\mathbf{R}_{\tempo}$ called $\tilde{\mathbf{R}}_{\tempo}$, where
\begin{align}\label{eq:var_ok_3}
&\left\{\tilde{\mathbf{R}}_{\tempo}\right\}_{n,n} = \frac{\sigma^2(1+\alpha)}{\lVert \mathbf{y}_{\risk} \rVert^2 / \Nrx},
\end{align} 
for $n \in \mathcal{I}_{\risk}$. In \eqref{eq:var_ok_3} $\lVert \mathbf y_{\risk} \rVert^2 / \Nrx$ is the average received power  measured at the receiver, and $\alpha$ is an empirical parameter introduced to oversize the variance of the noise in the observation model with respect to the expected value to account for the noise increase due to multipath.\footnote{This is the typical approach followed in the literature to account for potential model mismatch in the observation that might be caused, for instance,  by fading \cite{li2015resampling}.}

\section{\ac{RIS} and Precoder Optimization for User Tracking: Problem Formulation}
\label{sec:optimization} In this section, we will introduce a joint optimization problem for designing the transmission parameters, i.e., the precoding vectors and the \acp{RIS} reflection coefficients. At time $\tempo$ and for the $k$th \ac{RIS},  these parameters are grouped in a vector given by 
\begin{equation}\label{eq:optpara}
\txparamk^{(\tempo)}=(\risparamk^{(\tempo)}, \beta^{(\tempo)}_{\risk}),
\end{equation} 
where $\risparamk^{(\tempo)}$ contains the $k$th \ac{RIS} reflection coefficients and $\beta_k^{(\tempo)}$ are real-valued weights that determine the amount of power to be assigned to the $k$th \ac{RIS} -- under the constraint $\Sigma_{\risk = 1}^K  (\beta_{\risk}^{(\tempo)} )^2 = \powertx$, with $\powertx$ denoting the power budget allocated for the transmission from the \ac{BS} to the $K$ \acp{RIS}. In fact, as illustrated in Sec. \ref{BDP}, the precoding vector can be found using a \ac{BD} approach so that the overall precoding optimization problem turns out to be a power allocation problem to different \acp{RIS}, i.e., $\mathbf{f}_{\risk}^{(\tempo)} = \beta^{(\tempo)}_{\risk}\, \mathbf{v}^{({\mathrm{0}})}_{\risk}$.

The optimization of parameters in \eqref{eq:optpara} is performed during the time update stage of the EKF, where we predict the state and covariance for the next time step $\tempop$ with the availability of measurements up to time $\tempo$. 
Such measurements and information consist of $\mathbf{J}_{\tempop}$, $\bmkkp$, and $\boldsymbol{\Sigma}_{\tempop \lvert \tempo}$, among others {(see Alg. \ref{alg:EKF})}.   

\paragraph{Cost Function}

{To define a cost function for optimizing the transmission parameters at time $\tempo$, it is natural to consider the covariance matrix of the state at the next time $\mathbf{s}_{\tempo+1}$. Note that the expression of the covariance matrix $\boldsymbol{\Sigma}_{\tempop \lvert \tempop}$ of the Kalman filter used in Alg. 1 depends on $\mathbf{R}_{\tempop}$, which in turn depends on the position $\pRXtp$, not yet available. Therefore $\boldsymbol{\Sigma}_{\tempop \lvert \tempop}$ in Alg. 1 cannot be used for our purposes.}
{
Indeed, recalling \eqref{eq:noisem} and assuming the same empirical term $\alpha$ introduced in \eqref{eq:var_ok_3} to characterize the observation model mismatch, the covariance matrix of $\betakp$ is given by
\begin{align}\label{eq:Rn_n1}    
\left[\mathbb{E}\left\{ \boldsymbol{\eta}_{\tempop} \boldsymbol{\eta}^T_{\tempop}\right\} \right]_{n,n} & = \left[\Rkp\left(\sskp,{\txparam^{(\tempop)}}\right) \right]_{n,n} \\
& = {\bar{\sigma}_k^2} =\frac{\sigma^2(1+\alpha)}{P_{k}(\pRXtp,{\txparam^{(\tempop)}})}\nonumber
\end{align} 
with $n \in \mathcal{I}_k$, $\txparam^{(\tempop)} = \left\{\txparam_1^{(\tempop)},\ldots,\txparam_K^{(\tempop)}\right\}$, and $\powerk(\pRXtp,{\txparam^{(\tempop)}})$ representing the power received by the $k$-th \ac{RIS} when the node is at position $\pRXtp$ for a {determined} $\txparamk^{(\tempop)}$. It is also worth noting that the error covariance matrix in \eqref{eq:Rn_n1} does not depend on velocity, so we can simplify as $\Rkp(\sskp,{\txparam^{(\tempop)}}) =\Rkp(\pRXtp,{\txparam^{(\tempop)}})$.}

{To circumvent this problem, we derive a new cost function that can be evaluated at time $\tempo$, with the goal of minimizing the MSE of the \ac{UE} state estimate for the next time. To simplify the notation in the following, the temporal index $\tempop$ is omitted from the parameters $(\pRXtp,{\txparam^{(\tempop)}})$, i.e., $({\mathbf{p}},{\txparam})$ is used instead. 
 Let us now denote
\begin{align}\label{eq:S11}
& \Skp\left({\mathbf{p}},{\txparam}\right) = \Hkp \, \Pkpok\, \Hkp^T + \Rkp\left({\mathbf{p}},{\txparam}\right), \\
& \Kkp\left({\mathbf{p}},{\txparam}\right) = \Pkpok\, \Hkp^T \Skp^{-1}\left({\mathbf{p}},{\txparam}\right). \label{eq:K1}
\end{align}
}
{
{The state estimate at time $\tempop$ is}
\begin{equation}\label{eq-2}
\hsskp= \bmkkpp = \bmkkp + {\mathbf{K}}_{\tempop}\left({\mathbf{p}},{\txparam}\right) \, \innkp , 
\end{equation} 
with {the innovation vector given by}
\begin{align}\label{eq:observa_model}
&\innkp=\bzkp - \bhkkp = \Hkp \left(\sskp - \bmkkp\right) + \betakp.
\end{align} 
In the given expression $\bhkkp$ is the observation function evaluated in $\bmkkp$, i.e., $\bhkkp = h(\bmkkp)$, $\Hkp$ is the Jacobian matrix evaluated in $\bmkkp$ and $\bzkp$ is the observation signal in the next time instant position $\pRXtp$. Note that \eqref{eq:observa_model} is the linearized version of \eqref{eq:observationmodel} according to the first-order Taylor expansion. Under these conditions, we can condition on the future state $\sskp$ and introduce the cost function as follows
 \begin{align} \label{eq:costfunctionw0}
 \mathcal{E}\left({\mathbf{p}},{\txparam}\right) &= \text{Tr} \left( \mathbb{E}_{\boldsymbol{\eta}_{\tempop}} \left\{ \left( \sskp - \hsskp\right) \left(\sskp - \hsskp\right)^T\right\} \right) \nonumber \\
 &= \text{Tr}\left(\mathbb{E}_{\boldsymbol{\eta}_{\tempop}}
 \left\{\left(\Delta\sskp- {\mathbf{K}}_{\tempop}\left({\mathbf{p}},{\txparam}\right) \innkp\right) \cdot\nonumber \right.\right. \\ & \left.\left. \, ~ \times \left(\Delta\sskp- {\mathbf{K}}_{\tempop}\left({\mathbf{p}},{\txparam}\right) \innkp \right)^T\right\}\right),
\end{align} 
where $\text{Tr}\left(\cdot\right)$ is the trace operator, $\mathbb{E}_{\boldsymbol{\eta}_{\tempop}} \{ \cdot \}$ is the expected value evaluated with respect to the observation noise 
and $\Delta \mathbf{s}_{\tempop}=\left(\sskp - \bmkkp\right)$ is an unknown random variable depending on the future state.
}

{
From \eqref{eq:Rn_n1}-\eqref{eq:observa_model} we obtain the following
%
\begin{align}
    & \mathbb{E}_{\boldsymbol{\eta}_{\tempop}}\left\{\innkp\innkp^T\right\} = \Hkp\,\Delta\sskp\Delta\sskp^T\,\Hkp^T +\Rkp\left({\mathbf{p}},{\txparam}\right),\\
    & \mathbb{E}_{\boldsymbol{\eta}_{\tempop}} \left\{\innkp\Delta\sskp^T\right\} = \Hkp\,\left(\Delta\sskp\Delta\sskp^T\right),
\end{align} 
and it is easy to reformulate \eqref{eq:costfunctionw0} as 
\begin{align}
\label{eq:costfunctionwy}
\mathcal{E}\left({\mathbf{p}},{\txparam}\right)  =  \text{Tr} & \left(\Delta\sskp\Delta\sskp^T - {\mathbf{K}}_{\tempop}\left(\mathbf{p},{\txparam}\right) \Hkp \Delta\sskp\Delta\sskp^T \right.
\nonumber \\
 & \left. - \Delta\sskp\Delta\sskp^T \Hkp^T {\mathbf{K}}_{\tempop}\left(\mathbf{p},{\txparam}\right)^T\right.   \nonumber \\
 & \left. + {\mathbf{K}}_{\tempop}\left(\mathbf{p},{\txparam}\right) \left[ \Hkp\,\Delta\sskp\Delta\sskp^T\,\Hkp^T \right. \right.  \nonumber \\ &\left. \left.  + \Rkp\left({\mathbf{p}},{\txparam}\right)\right] {\mathbf{K}}_{\tempop}\left(\mathbf{p},{\txparam}\right)\right)  .
\end{align}
}

{
\paragraph{Problem Formulation}
We are now in a position to define the optimization problem at hand, i.e.:
\begin{equation}\label{eq:problem11111}
\begin{array}{c}
\underset{\substack{{\txparam}}} 
 {\min} ~~ \mathbb{E}_{\sskp \lvert \tempo}\left\{\mathcal{E}\left({\mathbf{p}},{\txparam}\right)\right\}\\
\text{s.t.}
\quad {{\txparam}} \in {\mathcal{Q}}  \, 
\end{array}, 
\end{equation} 
where $\mathbb{E}_{\sskp \lvert \tempo}$ is the average over $\mathbf{s}_{\tempop \lvert \tempo} \sim \mathcal{N}\left(\bmkkp,\boldsymbol{\Sigma}_{\tempop \lvert \tempo}\right)$ and $\mathcal{Q}$ includes the precoder and \acp{RIS} constraints, i.e., $\sum_{k=1}^K \beta_k^2 \le \powertx$, $\lvert \tkp \lvert \le 1, \, \forall  \, k \in \mathcal{K} \,, p \in \mathcal{P}$.
Since $\mathbb{E}_{\sskp \lvert \tempo}\left\{\Delta \sskp \Delta \sskp^T\right\} = \boldsymbol{\Sigma}_{\tempop \lvert \tempo}$, it is easy to verify that if we replace $\mathbf{p}$ in \eqref{eq:costfunctionwy} with the actual UE position, then $\mathbb{E}_{\sskp \lvert \tempo}\left\{\mathcal{E}\left({\mathbf{p}},{\txparam}\right)\right\} = \text{Tr} \left(\boldsymbol{\Sigma}_{\tempop \lvert \tempop}\right)$ \cite{anderson2012optimal}.
Thus, the problem considered consists in minimizing the average of the trace of the covariance matrix of $\mathbf{s}_{\tempop}$, based on observations at time $\tempo$.}
{However, the computation of the expected value with respect to the position in \eqref{eq:costfunctionwy} is very complex, given the inverse proportionality of $\Rkp({\mathbf{p}},{\txparam})$ in $\Kkp({\mathbf{p}},{\txparam})$. As such, in order to reduce the computational complexity of the problem, the idea could be to introduce a fixed Kalman gain matrix $\tilde{\mathbf{K}}_{\tempop}$ that does not depend on the error covariance matrix $\Rkp({\mathbf{p}},{\txparam})$ but is expressed in terms of an approximation for the covariance matrix, fixed and independent of the transmission parameters.
Accordingly we consider $\tilde{\mathbf{K}}_{\tempop}$ to be used in \eqref{eq:problem11111} as
\begin{equation}\label{eq_Ktilde}
 \tilde{\mathbf{K}}_{\tempop} = \Pkpok \Hkp^T\left(\Hkp \, \Pkpok \, \Hkp^T+\breve{\mathbf{R}}_{\tempo} \right)^{-1} 
 \end{equation} 
where $\breve{\mathbf{R}}_{\tempo}$ is an estimate at time $\tempo$ of the noise covariance matrix at time $\tempo + 1$. To evaluate $\breve{\mathbf{R}}_{\tempo}$, the working principle of our approach is exploited for the optimization algorithm. Due to the power balancing effect introduced by the proposed approach (see next section), we expect that the power distribution between two successive steps is not very different and so is the \ac{SNR}. Therefore, the noise covariance matrix $\Rkp$, which is inversely proportional to the
\ac{SNR}, is not significantly different from the $\mathbf{R}_{\tempo}$ at the previous time points. Therefore, to exploit this correlation, it is reasonable to use the covariance matrix estimated by \eqref{eq:var_ok_3}. However, \eqref{eq:var_ok_3} depends on the power received at time $\tempo$, which is affected by the power allocation terms $\beta_k^{(\tempo)}$ evaluated in the previous time step, so directly using \eqref{eq:var_ok_3} to evaluate $\tilde{\mathbf{K}}_{\tempop}$ would transfer the effect of the previous optimization to the next. Accordingly, we take:
\begin{align}\label{eq:var_ok_3bis}
&\left\{\breve{\mathbf{R}}_{\tempo}\right\}_{n,n} = \frac{\left[\beta_k^{(\tempo)}\right]^2}{P_{\text{tx}}/K} \cdot \frac{\sigma^2(1+\alpha)}{\lVert \mathbf{y}_k \rVert^2 / \Nrx},
\end{align} 
for $n \in \mathcal{I}_k$. Note that $\breve{\mathbf{R}}_{\tempo}$ is an estimate of the noise covariance matrix when all \acp{RIS} are assigned the same power $P_{\text{tx}}/K$, and as such it is independent of the prior power allocation strategy. In what follows, for notational convenience, we omit the dependence of all quantities involved on time $\tempop$ and $\tempo$ unless explicitly required.
Starting from \eqref{eq:problem11111} and considering only the terms depending on the optimization variable $\txparam$, the optimal parameters of the transmission configuration are obtained by solving the problem:
\begin{equation}\label{eq:problem1111}
\begin{array}{c}
\underset{\substack{{\txparam}}} 
 {\min} ~~ \text{Tr} \left(\tilde{\mathbf{K}}  ~ \mathbb{E}_{\mathbf{p}} \left\{\mathbf{R}\left(\mathbf{p},{\txparam}\right)\right\} \tilde{\mathbf{K}}^T \right)\\
\text{s.t.}
\quad {{\txparam}} \in {\mathcal{Q}}  \, 
\end{array}, 
\end{equation} 
where {$\mathbb{E}_{\mathbf{p}} \triangleq \mathbb{E}_{\mathbf{p}_{\tempop}\lvert {\tempo}}$ is the expectation with respect to the position only, being the quantities only depending on $\mathbf{p}$ }. 
The proof of the equivalence between \eqref{eq:problem11111} and \eqref{eq:problem1111} can be found in the supplementary material in {Appendix B}. 
}
Let us denote by $\mathbf{u}_{n}$ the $n$th column of $\tilde{\mathbf{K}}$. Thus, by introducing $\xi_{ \risk} = \sigma^2(1+\alpha)\sum\limits_{n\in \mathcal{I}_k}\rVert \mathbf{u}_{n}\rVert^2$, the problem can be further reformulated as 
\begin{equation}\label{eq:wMMSE1_21}
\begin{array}{c}
\underset{\substack{{\txparam}}} 
 {\min} ~~ \Sigma_{k=1}^{K} \xi_{\risk} \,{\mathbb{E}_{ \mathbf{p} } \left \{ \right.} \left({\powerk({\mathbf{p}},{\txparamk})}\right)^{-1}  {\left.\right\}}\\
\text{s.t.}
\quad {{\txparam}} \in {\mathcal{Q}}  \, 
\end{array} .
\end{equation} 
Let us now introduce $P^{(0)}_{k}({\mathbf{p}},\risparamk)$ as the received power corresponding to unitary power distribution. Consequently, the actual received power is  $\powerk({\mathbf{p}},\txparamk) = \left[\beta_k\right]^2 \, P^{(0)}_{k}({\mathbf{p}},\risparamk)$ and we can recast problem \eqref{eq:wMMSE1_21} as 
\begin{equation}\label{eq:wMMSE1_211}
\begin{array}{c}
\underset{\substack{{\txparam}}} 
 {\min} ~~ \sum_{k=1}^{K}\xi_{\risk} \frac{1}{\left[\beta_k\right]^2} \mathbb{E}_{ \mathbf{p} }\left\{ P^{(0)}_{k}({\mathbf{p}},\risparamk) ^{-1}\right\}\\
\text{s.t.}
\quad {{\txparam}} \in {\mathcal{Q}}  \, 
\end{array} .
\end{equation} 
Since the $k$th \ac{RIS} design concerns only the term $\mathbb{E}_{ \mathbf{p}}\left\{ P^{(0)}_{k}({\mathbf{p}},\risparamk) ^{-1}\right\}$, the design criterion for the \acp{RIS} becomes  directly {a SNR optimization criterion considering the possible UEs positions}, i.e. 
\begin{equation}\label{eq:wMMSE1_2111}
\begin{array}{c}
\underset{\substack{{\risparamk}}} 
 {\min} ~~  \mathbb{E}_{ \mathbf{p}}\left\{ P^{(0)}_{k}({\mathbf{p}},\risparamk) ^{-1}\right\}\\
\text{s.t.}
\quad |\tkp| \le 1, \, \forall p \in \mathcal{P}
\end{array}. 
\end{equation} 
Let us assume that problem \eqref{eq:wMMSE1_2111} is solved and let us denote by $\risparam^*_k$ and $\pi_k$, respectively, the optimal $\risparam_k$ and the value of the objective function, i.e., 
\begin{equation}\label{eq_gk}
\pi_k = \mathbb{E}_{ \mathbf{p}}\left\{ P^{(0)}_{k}({\mathbf{p}},\risparam^*_k) ^{-1}\right\}.
\end{equation} Hence, the precoder optimization problem can be written 
{in the form of a power splitting problem among different RISs, i.e., 
 }\begin{equation}\label{eq:wMMSE1_21111}
\begin{array}{c}
\underset{\substack{{\beta_k}}} 
 {\min} ~~  \sum_{k=1}^{K} \gamma_{k} \frac{1}{\beta_k^2}\\
\text{s.t.}
\quad \Sigma_{k = 1}^K \left[ \beta_k\right]^2 \le \powertx
\end{array} 
\end{equation} 
where $\gamma_{ k} = \xi_{ \risk} \, \pi_k$.

\section{\ac{RIS} and Precoder Optimization for User Tracking: Proposed Solutions}
\label{sec:soloptimization}

\subsection{\ac{RIS} {O}ptimization}\label{RIsopt}
Let us first focus on problem \eqref{eq:wMMSE1_2111}. From the assumption that each \ac{UE} antenna receives the same power, the power of the signal coming from the $k$-th \ac{RIS} can be evaluated as the power coming from the first \ac{UE} antenna, i.e.:
\begin{align}
\powerk^{(0)}({\mathbf{p}},\risparamk)
&= (\mathbf{v}_k^{(0)})^H \mathbf{G}_k^H \RISPARAMk^H \mathbf{b}_k^H(\mathbf{p}) \mathbf{b}_k(\mathbf{p})\RISPARAMk\mathbf{G}_k \mathbf{v}_k^{(0)}
\label{eq:Power_def}
\end{align} 
where $\mathbf{b}_k(\mathbf{p}) \in \mathbb{C}^{1 \times P}$ is the first row of $\mathbf{B}_k(\mathbf{p})$ corresponding to the first receiving antenna, and we recall that $\RISPARAMk= \operatorname{diag} \left(\risparamk\right)$ is the matrix containing the \ac{RIS} coefficients vector. Without loss of generality, the index $k$ is neglected from now on, and a formulation valid for every \ac{RIS} is obtained. The goal of the \ac{RIS} optimization in \eqref{eq:wMMSE1_2111} is to find the beamforming vector $\risparam$ such that the (statistical) average of the inverses of \eqref{eq:Power_def} over the uncertainty range of $\mathbf{p}$ is minimized. Consequently, we can formulate \eqref{eq:wMMSE1_2111}  as \begin{equation}\label{P_min_in_avg_power1}
\begin{array}{c}
\underset{\substack{{\risparam}}} 
 {\min} ~~  \int {P}^{(0)}\left(\text{\boldmath{$p$}},\risparam\right)^{-1} {f_{{\mathbf{p}}}(\text{\boldmath{$p$}})} d\text{\boldmath{$p$}}\\
\text{s.t.}
\quad |\risparamel_p|^2 \le 1 \, ~ \forall p \in \mathcal{P}
\end{array} 
\end{equation} 
where {$f_{{\mathbf{p}}}(\text{\boldmath{$p$}})$} is the a-priori \ac{pdf} of the position {$\text{\boldmath{$p$}}$}. 
In particular, from the \ac{EKF} model, we have {$f_{{\mathbf{p}}}(\text{\boldmath{$p$}}) = \mathcal{N}({\mathbf{p}};{\mathbf{m}_{\tempop|\tempo}^{\{1:3\}},{\boldsymbol{\Sigma}}_{\tempop|\tempo}^{\{1:3,1:3\}}})$}. \footnote{{The notations $\mathbf{m}_{\tempop|\tempo}^{\{1:3\}}$ and ${\boldsymbol{\Sigma}}_{\tempop|\tempo}^{\{1:3,1:3\}}$ indicate respectively the first three elements of $\mathbf{m}_{\tempop|\tempo}$ and the first $3 \times 3$ block of ${\boldsymbol{\Sigma}}_{\tempop|\tempo}$. }} 
The problem \eqref{P_min_in_avg_power1} is nonconvex and therefore, we propose below an iterative \ac{BCD} algorithm to find a local optimum. 
\paragraph{\ac{BCD} Algorithm} For the sake of notation, we set $\mathbf{g} \triangleq \mathbf{G}{\mathbf{v}^{(0)}}$ and we consider the following equivalence \begin{align}
\label{eq:equivalence}
&\mathbf{b}(\mathbf{p})\mathbf{C}\mathbf{g} = \mathbf{b}(\mathbf{p}) \text{diag}\left({\mathbf{g}}\right)\risparam = \tilde{\mathbf{h}}(\mathbf{p}) \risparam, 
\end{align} where $\tilde{\mathbf{h}}(\mathbf{p}) \triangleq \mathbf{b}(\mathbf{p}) \text{diag}\left({\mathbf{g}}\right) \in \mathbb{C}^{{1} \times P}$. Accordingly, \eqref{eq:Power_def} simplifies as \begin{align}
\label{P:power_def}
{P}^{(0)}\left(\mathbf{p},\risparam\right) = \risparam^H \tilde{\mathbf{h}}^H(\mathbf{p})\tilde{\mathbf{h}}(\mathbf{p}) \risparam.
\end{align} 
Let us introduce the following term:
\begin{align}
\label{eq:mMSEMat0_21}
{\Sigma}\left(\mathbf{p},\risparam\right) & = \frac{\delta}{{P}^{(0)}\left(\mathbf{p},\risparam\right) + \delta} \, .
\end{align}
When $\delta$ is small, we can recast problem in \eqref{P_min_in_avg_power1} as 
\begin{equation}\label{Emin_problem}
\begin{array}{c}
\underset{\substack{{\risparam}}} 
 {\min} ~~  \int {\Sigma}\left(\text{\boldmath{$p$}},\risparam\right) f_{{\mathbf{p}}}(\text{\boldmath{$p$}}) d\text{\boldmath{$p$}} \\
\text{s.t.}
\quad |\risparamel_p|^2 \le 1, \, ~ \forall p \in \mathcal{P}
\end{array} 
\end{equation} 
Then, we introduce
\begin{align}
\label{eq:mMSEMat0_22}
{\Upsilon}\left(\mathbf{p},\risparam,w\right) & = 1 + |w|^2 {P}^{(0)}\left(\mathbf{p},\risparam\right) - 2 \Re \{ w \tilde{\mathbf{h}}(\mathbf{p}) \risparam \} + \delta |w|^2,
\end{align} 
where $w$ is a complex parameter. It is straightforward to get the equivalence
\begin{equation}\label{Equivalence}
\begin{array}{c}
{\Sigma}\left(\text{\boldmath{$p$}},\risparam\right) = \min \limits_{w} {\Upsilon}\left(\text{\boldmath{$p$}},\risparam,w\right) = {\Upsilon}\left(\text{\boldmath{$p$}},\risparam,w^*\right),
\end{array} 
\end{equation}
where
\begin{equation}
\label{eq:wott}
w^* = \frac{ \risparam^H \tilde{\mathbf{h}}^H(\mathbf{p}) }{{{P}^{(0)}\left(\mathbf{p},\risparam\right)}+\delta} \, . 
\end{equation}
Accordingly, the problem in \eqref{P_min_in_avg_power1} can be reformulated as 
\begin{equation}\label{Emin_problem11}
\begin{array}{c}
\underset{\substack{{\risparam}}} 
 {\min} ~~  \int \min \limits_{w}\left[ {\Upsilon}\left(\text{\boldmath{$p$}},\risparam,w\right)\right] f_{{\mathbf{p}}}(\text{\boldmath{$p$}}) d\text{\boldmath{$p$}} \\
\text{s.t.}
\quad |\risparamel_p|^2 \le 1, \, ~ \forall p \in \mathcal{P}.
\end{array} 
\end{equation}

Note that when $\delta$ tends to zero, ${\Sigma}\left(\text{\boldmath{$p$}},\risparam\right)$ 
 in \eqref{Equivalence} also tends to zero for each $\mathbf{c}$ and $\mathbf{p}$ and the problem \ref{Emin_problem11} cannot provide a meaningful solution. Therefore, $\delta$ must be chosen appropriately so that the problem \eqref{Emin_problem11} is not starved in the zero solution, and, at the same time $\delta \ll  {P}^{(0)}\left(\mathbf{p},\risparam\right)$ to ensure equivalence between \eqref{P_min_in_avg_power1} and \eqref{Emin_problem}. This aspect is explained in more detail below.
 
 Although \eqref{Emin_problem11} is not convex, it is convex for the remaining variable if one of the two variables is fixed. As a result, a good local optimum can be obtained using the iterative \ac{BCD} approach \cite{Bertsekas1999}, where all variables involved are updated sequentially at each iteration. To illustrate this, we denote by $\risparam^{(q)}$ the value of $\risparam$ after the $q$th iteration of \ac{BCD}. From \eqref{eq:wott} we can then compute ${w}^{(q)}$ as follows 
 \begin{align}
\label{eq:wott_it}
& w^{(q)}\left(\mathbf{p}\right) = \frac{(\risparam^{(q)})^H \tilde{\mathbf{h}}^H(\mathbf{p})}{{{P}^{(0)}\left(\mathbf{p},\risparam^{(q)}\right)}+\delta^{(q)}},
\end{align} 
where
\begin{align}
\label{eq:delta_q}
&\delta^{(q)} =  \frac{\min_{\mathbf{p}} P^{(0)}\left(\mathbf{p},\risparam^{(q)}\right)}{M}, 
\end{align}
where $M \gg 1$. The optimal choice of $M$ depends on the trade-off between fast convergence ($M$ must not be excessively high) and equivalence between \eqref{P_min_in_avg_power1} and \eqref{Emin_problem} ($M$ must be high). Finding an analytical approach to optimally deal with this trade-off is a very complex task. However, based on numerical considerations, we found that a value of $M = 100$ is a good compromise, and therefore we set this value in the numerical results section. In \eqref{P_min_in_avg_power1} and \eqref{Emin_problem}
the problem of finding the optimal $\risparam^{(q+1)}$ for a given ${w}^{(q)}$ can be solved as follows. Let us first introduce the terms
\begin{align}
\label{Emin_problem1bis}
&\mathbf{A}^{(q)} = \int |w^{(q)}\left(\text{\boldmath{$p$}}\right)|^2\tilde{\mathbf{h}}^H(\text{\boldmath{$p$}})\tilde{\mathbf{h}}(\text{\boldmath{$p$}}) f_{{\mathbf{p}}}(\text{\boldmath{$p$}})d\text{\boldmath{$p$}} && \in \mathbb{C}^{{P} \times P} \\\nonumber
& \mathbf{s}^{(q)} = \int w^{(q)}\left(\text{\boldmath{$p$}}\right) \tilde{\mathbf{h}}(\text{\boldmath{$p$}}) f_{{\mathbf{p}}}(\text{\boldmath{$p$}})d\text{\boldmath{$p$}} \in \mathbb{C}^{{1} \times P} && \in \mathbb{C}^{{1} \times P} 
\end{align} 
which can be computed by numerical integration. Therefore, the optimal $\risparam$ for a given $w^{(q)}(\mathbf{p})$ can be found from \begin{equation}\label{Emin_problem2}
\begin{array}{c}
\underset{\substack{{\risparam}}} 
 {\min} ~~  \risparam^H \mathbf{A}^{(q)} \risparam -2 \Re \{\mathbf{s}^{(q)} \risparam\} \\
\text{s.t.}
\quad | \risparamel_p | ^2\le 1, \, \forall p \in \mathcal{P}.
\end{array} 
\end{equation} This is a convex problem that can be solved using traditional numerical approaches whose solution will be denoted as OPT in the sequel. However, the inclusion of a large number of inequality constraints makes the problem computationally difficult when $P$ is high. An iterative strategy based on \ac{AO}, where optimization is performed at each step for a single element $\risparamel_p$ while the others remain fixed, is one way to simplify the problem. The \ac{AO} approach will be referred to as OPT-AO. We will show in the results  that OPT-AO achieves performance very close to that of OPT, while noticeably reducing computational complexity.
To elaborate, consider the following problem 
\begin{equation}\label{Emin_problem3}
\begin{array}{c}
\underset{\substack{{\risparamel_p }}} 
 {\min} ~~  \risparam^H \mathbf{A}^{(q)} \risparam -2 \Re \{\mathbf{s}^{(q)} \risparam\} \\
\text{s.t.}
\quad | \risparamel_p | ^2\le 1 \,, 
\end{array} 
\end{equation}
where the difference with respect to \eqref{Emin_problem2} is that the minimum is searched for each $\risparamel_\riselem$ independently. The solution of this problem can be found in closed form by enforcing the \ac{KKT} optimality conditions. In particular, the Lagrangian function is \begin{align}
\label{eq:Lag_fun}
&L\left(\risparamel_p,{\lambda}_p\right) = \risparam^H \mathbf{A}^{(q)} \risparam -2 \Re \{\mathbf{s}^{(q)} \risparam\} + \lambda_p \left( |\risparamel_p | ^2 - 1\right).
\end{align} Considering the KKT conditions we have \begin{align}
\label{eq:KKTcond}
&\nabla_{\risparamel_p} L\left(\risparamel_p,{\lambda}_p\right) = 2\mathbf{a}^{(q)}_p \risparam - 2 {\left(s^{(q)}_p\right)^*} + 2 \lambda_p \risparamel_p = 0\\
&\lambda_p \ge 0\\
&\lambda_p \left( |\risparamel_p | ^2 - 1\right) = 0,
\end{align} where $\mathbf{a}^{(q)}_p \in \mathbb{C}^{{1} \times P}$ is the $p$th row of $\mathbf{A}^{(q)}$, $s^{(q)}_p$ is the $p$th element of $\mathbf{s}^{(q)} $ and the second condition is the complementary slackness condition. {Denote by $V_p = \Sigma_{l \ne p} a^{(q)}_{p,l} \risparamel_l$}, $A_p = a^{(q)}_{p,p}$, and $Y_p = (s^{(q)}_p)^*-V_p$, where $a^{(q)}_{p,l}$ is the $l$th element of $\mathbf{a}^{(q)}_p$, it is then easy to derive from \eqref{eq:KKTcond}
\begin{align}
\label{eq:solution}
&\risparamel_p = {Y_p}/{A_p}, ~\lambda_p = 0 & \text{if } |Y_p| < |A_p|\\ \nonumber
&\risparamel_p = {Y_p}/{|Y_p|}, ~\lambda_p = |Y_p|-A_p & \text{if } |Y_p| \ge |A_p|.
\end{align} 

\begin{algorithm}[t!]
\SetAlgoLined
{\textbf{Uncertainty area construction}};\\
\For{$q=1, \ldots, N_{r}$}{
Compute $f^{(q)}_{{\mathbf{p}}}(\boldsymbol{p})$ according to \eqref{Model_cov_joint}-\eqref{joint_pdf}; \\
}
$f_{{\mathbf{p}}}(\boldsymbol{p})= \frac{1}{N_{r}}\Sigma_{q=1}^{N_{r}} f^{(q)}_{{\mathbf{p}}}(\boldsymbol{p})$;\\
{\textbf{\ac{RIS} Parameter Update}};\\
Initialize $\risparam^{*}=\risparamk^{(\tempo)}$;\\
\For{$q=1, \ldots, N_{\text{BCD}}$} {
Initialize the iterative solution: $\risparam^{(q)}=\risparam^*$;\\
Update $w^{(q)}$, $\textbf{A}^{(q)}$ and $\textbf{s}^{(q)}$ according to \eqref{eq:wott_it}-\eqref{Emin_problem1bis};\\
\For{p=1, \ldots , P}
{Solve problem \eqref{Emin_problem3} for $\risparamel_p$ as in \eqref{eq:solution};
}
}
$\risparam_{k}^{(\tempop)}=\risparam^{*}$
\caption{\ac{RIS} Optimization with BCD Approach}
\label{alg:RIS_Opt}
\end{algorithm} 
%
\subsection{Precoder Optimization}
The precoder optimization problem presented in \eqref{eq:wMMSE1_21111} can be easily solved by resorting to the \ac{KKT} constraints. Writing the Lagrangian function as 
\begin{equation}
 L(\beta_k,\lambda)= \Sigma_{k=1}^K \frac{\gamma_{k}}{\beta_k^2}+\lambda \left (\sum_{k=1}^K \beta_k^2 - \powertx \right),
\end{equation} 
the \ac{KKT} conditions can be derived as 
\begin{align}
 & \nabla_{\beta_k} L(\beta_k,\lambda)= - 2 \frac{\gamma_{k}}{\left(\beta^*_k\right)^3} + 2 \lambda^* \beta_k^*=0\\
& \Sigma_{k=1}^K {\left(\beta^*_k\right)^2} - \powertx \leq 0\\
&\lambda^* \geq 0\\
&\lambda^*\left(\Sigma_{k=1}^K {\left(\beta^*_k\right)^2} - P_{\text{tx}}\right) =0 \, .
\end{align} 
The existence conditions of the given equations are summarized as $\beta_k^* \neq 0$ and $\lambda^* \neq 0$, and thus, it is easy to derive the optimal solution as \begin{align} \label{eq:beta_and_lambda}
    & \lambda^*= \frac{\left(\sum_{k=1}^K \sqrt{\gamma_{k}}\right)^2}{\powertx^2}, && \left(\beta^*_k \right)^2 = \frac{\sqrt{\gamma_{k}}\, \powertx}{{\sum_{k=1}^K \sqrt{\gamma_{k}}}} \, .
\end{align}

{The proposed power allocation strategy at each time $\tempo$ is not simply a matter of optimizing the SNR of each connection, but a much broader problem, similar to waterfilling where most of the power is distributed in favor of the channels with the highest SNR to maximize the achievable communication rate. Instead, in our case the power weights depend, among other factors, on the Kalman gains defined in \eqref{eq_Ktilde}, which in turn depend on the Jacobian matrices evaluated considering the near-field propagation model between the RIS and the UE. Therefore, the near-field model is exploited at each stage of the proposed strategy.}

\subsection{Two-{T}imescale {O}ptimization }

\begin{figure}
    \centering
    \resizebox{\columnwidth}{!}
    {\input{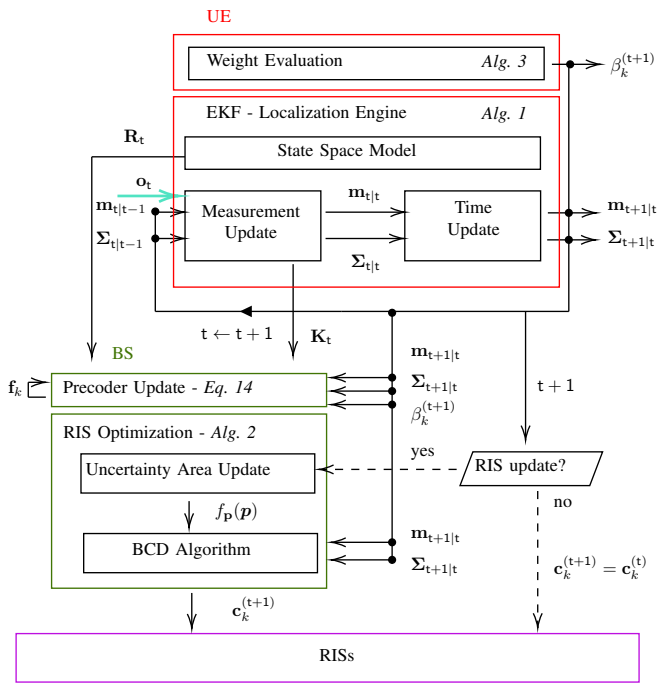}}
    \caption{Interconnections between localization, RIS optimization, and BS precoding algorithms.}
    \label{fig:alg}
\end{figure}

As discussed in Sec. \ref{Main_Contributions}, \ac{RIS} and precoding optimization are performed on two different time scales. As for the precoding, the time scale corresponds to the interval between two consecutive pilot signal transmissions and may be on the order of a few milliseconds, e.g., on the same time scale as channel estimation in communications. This interval corresponds to the time between two successive location updates of the EKF algorithm and will be denoted by $d\tempo$ 
in the following. Conversely, the time scale of the \ac{RIS} update, denoted by $T_O$, can be expected to be on the order of seconds, i.e., orders of magnitude higher. We denote with $N_{r} = T_O/ d\tempo$ the (integer) number of steps between two successive \ac{RIS} optimizations. 
 Thus, if the previous \ac{RIS} optimization is performed at time $\tempo$, in order to obtain the new optimization at time $\tempo+N_{r}$ according to the approach presented in Section \ref{RIsopt}, the a-priori {\ac{pdf}} $f_{{\mathbf{p}}}(\boldsymbol{p})$ must be defined to account for the $N_{r}$ transitions of the EFK. More precisely, we have 
\begin{align}\label{Model_cov_joint}
 & \mathbf{m}_{\tempo+q+1|\tempo} = \A\mathbf{m}_{\tempo+q|\tempo},
 && \boldsymbol{\Sigma}_{\tempo+q+1|\tempo} = \A \boldsymbol{\Sigma}_{\tempo+q|\tempo} \A^T+\Q,
\end{align} 
for $q = 1,\ldots,N_{r}$.
The pdf of the position $\boldsymbol{p}$ at time $\tempo+q$ is given by {$f^{(q)}_{{\mathbf{p}}}(\boldsymbol{p}) = \mathcal{N}(\mathbf{m}_{\tempo+q+1|\tempo},\boldsymbol{\Sigma}_{\tempo+q+1|\tempo})$} and 
\begin{equation}\label{joint_pdf}
 \begin{aligned} f_{{\mathbf{p}}}(\boldsymbol{p}) = \frac{1}{N_{r}}\Sigma_{q=1}^{N_{r}} f^{(q)}_{{\mathbf{p}}}(\boldsymbol{p}) 
 \end{aligned} \, .
\end{equation} 
The terms in \eqref{Emin_problem1bis} are then evaluated by resorting to  Monte Carlo numerical integration by drawing samples according to \eqref{joint_pdf}. 
The algorithms for \ac{RIS} and precoder  optimization are sketched in Algorithm \ref{alg:RIS_Opt}, where $N_{\text{BCD}}$ indicates the number of iterations of the \ac{BCD}, and Algorithm \ref{alg:RIS_and_Precoder_Opt}.
{Fig. \ref{fig:alg} schematically represents the whole procedure entailing joint RIS and precoder optimization with UE tracking}.

\begin{algorithm}[t!]
\SetAlgoLined
 \For{${\risk}=1, \ldots, {K}$}{
\uIf{$N_{r} | \tempo$} 
{\text{Optimize the RIS};\\
}
\Else{{$\risparamk^{(\tempop)}=\risparamk^{(\tempo)}$}}
Evaluate $\tilde{\mathbf{K}}$ (see eq. \eqref{eq_Ktilde}), $\mathbf{u}_{n}$ and compute $\mathcal{I}_{\risk}$; \\
$\xi_{\risk}= \sigma_{\risk}^2 \, \Sigma_{n \in \mathcal{I}_{\risk}} \, \lVert \textbf{u}_n \rVert ^2$;\\
    $\risparamk^*=\risparamk^{(\tempop)}$;\\ 
Compute $\pi_{\risk}$ according to \eqref{eq_gk};\\
$\gamma_{\risk}=\xi_{\risk} \, \pi_{\risk}$;\\
$\left[\beta_{\risk}^{(\tempop)}\right]^2=\frac{\sqrt{\gamma_{\risk}}\,\powertx}{\Sigma_{\risk=1}^{{K}} \sqrt{\gamma_{\risk}}}$;\\
}
\caption{Optimization for Target Tracking }
\label{alg:RIS_and_Precoder_Opt}
\end{algorithm}
\section{Results}
\label{sec:results}
\begin{figure}
\centering
\psfrag{a}[lc][lc][0.8]{1 m}
\psfrag{b}[lc][lc][0.8]{2 m}
\psfrag{c}[lc][lc][0.8]{3 m}
\psfrag{k}[lc][lc][0.8]{20 m}
\psfrag{e}[lc][lc][0.8]{30 m}
\psfrag{f}[lc][lc][0.8]{30 m}
\psfrag{g}[lc][lc][0.8]{30 m}
\psfrag{h}[lc][lc][0.8]{BS}
\psfrag{i}[lc][lc][0.8]{RIS}
\psfrag{j}[lc][lc][0.8]{UE}
\psfrag{x}[lc][lc][0.7]{\textit{x}}
\psfrag{y}[lc][lc][0.7]{\textit{y}}
\psfrag{z}[lc][lc][0.7]{\textit{z}}
\includegraphics[width=1\linewidth,draft=false]{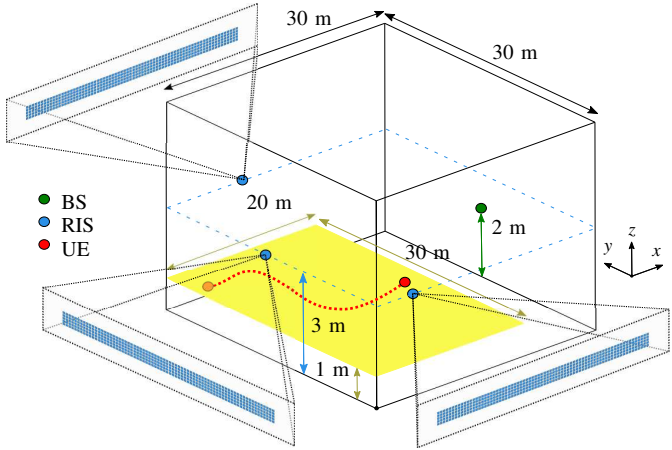}
\caption{Working scenario with an example of \ac{UE} trajectory. }
\label{fig:scenario_sim}
\end{figure} 
This simulation study aims to evaluate the validity of the optimized \ac{RIS}-based \ac{UE} tracking approach in a typical indoor scenario according to the following approaches: 
\begin{itemize}
    \item The optimization approach described in Sec.~\ref{RIsopt} with uniform power allocation for the different \acp{RIS} \footnote{{
    Uniform power allocation means that the joint RIS-BS precoder optimization results in
BD precoding with uniform power and RIS optimization.}}, i.e., $\beta^2_k = P_{\text{tx}}/K$, referred as OPT and its approximation obtained using the \ac{AO} method (Alg. \ref{alg:RIS_Opt}), referred as OPT-AO.
    \item The optimization approach described in Sec.~ \ref{RIsopt} with optimal power allocation denoted by $\beta$OPT and its \ac{AO} version (Alg. \ref{alg:RIS_Opt}) denoted as $\beta$OPT-AO. {In both cases, power allocation (\ac{BS} precoding) is performed according to Alg. \ref{alg:RIS_and_Precoder_Opt}.}
    \item The FOCUS and $\beta$FOCUS approaches, in which the \ac{RIS} design is performed to maximize the reflected power in the direction of the position estimate, as proposed in \cite{boyu2022bayesian}, while the power is allocated uniformly (FOCUS) or according to the Alg. \ref{alg:RIS_and_Precoder_Opt} ($\beta$FOCUS).
    \item The ROPT approach, in which the \ac{RIS} design is performed with the goal of maximizing the average communication rate when a-priori statistical knowledge about the \ac{UE} location is available and given by $f_{{\mathbf{p}}}(\mathbf{p})$. This approach has been proposed in \cite{abrardo2021intelligent}, and power is uniformly allocated. {We use this algorithm to investigate the loss of tracking performance occurring when \acp{RIS} and precoders are not optimized for localization purposes, and to study the trade-off that occurs between rate and position error when communication or localization objective functions are considered.}
\end{itemize}

\begin{table}[t]
\begin{center}
\makebox[\linewidth]{
\begin{tabular}{ |c|c| } 
 \hline
 \rowcolor{silver} PARAMETER & VALUE  \\
  \hline
     Carrier frequency & $f_c= 28 \, \mathrm{GHz}$  \\ \hline
     Subcarrier Bandwidth & $B_c= 120 \, \mathrm{kHz}$  \\ \hline
     Transmit power & $\powertx= 23 \, \mathrm{dBm}$  \\ \hline
      Noise power density & $NPD= -174 \, \mathrm{dB/Hz}$  \\ \hline
       Noise Figure & $NF= 7 \, \mathrm{dB}$  \\ \hline
 Rice Factor & $\kappa_b=5$ \\ \hline
 Empirical parameter for \eqref{eq:var_ok_3} and \eqref{eq:var_ok_3bis} &  $\alpha=0.5$ \\ \hline

 \end{tabular}
}
\end{center}
\caption{{System parameters for the simulation configuration.}}
\label{tab:sys_param}
\end{table}

\paragraph{Complexity Analysis}
{The two-time scale optimization approach allows us to consider the complexity problem as divided into two parts. During the shorter time scale operations, i.e., the location update time of $d\tempo$ seconds, the complexity of all algorithms is the same and is equal to the complexity of the standard \ac{EKF} for tracking. In this phase, the complexity is dominated by the computation of the inverse of the covariance matrix of the innovation, which is $\mathcal{O}\left(K^3\, \Nrx^3\right)$, where $K$ is the number of RIS and $\Nrx$ is the number of receiving antennas.}
{On the other hand, the \ac{RIS} design is clearly the bottleneck in terms of computational complexity since it involves solving a high-dimensional optimization problem whose dimension depends on the large number of RIS elements. In this respect, the considered approach OPT-AO allows a significant complexity reduction compared to traditional RIS optimization schemes, denoted as ROPT. More precisely, the complexity of the proposed algorithm OPT-AO is $\mathcal{O}(P)$, i.e., it is linear with the number of RIS elements (see Eq. \eqref{eq:solution}), while the complexity of the rate optimization scheme proposed in
 \cite{abrardo2021intelligent} is $\mathcal{O}(P^3)$. Accordingly, the complexity of OPT-AO is comparable to that of the simple FOCUS approach, which is also linear with $P$. However, it is worth noting that RIS optimization in the proposed scheme is performed on a long time scale (once every $T_O$ second), and therefore the complexity issues are significantly mitigated.}

\paragraph{Parameter Settings} We refer to the 3GPP specifications for 5G localization in an indoor open office (IOO) scenarios and the corresponding results reported in \cite{5Gloc_IEEE}. {Some of the system parameters are reported in Tab. \ref{tab:sys_param} with some of them} 
adjusted to match the narrowband \ac{RIS}-based near-field localization scenario considered in this work. Specifically, in the proposed scenario, localization is performed over a single $120$ $\mathrm{kHz}$ subcarrier band with a correspondent pilot symbol time $T_s = 8.3$ $\mathrm{\mu}$s. The transmitted pilot signal consists of $L = 100$ symbols, corresponding to a time of $\tau = 100\cdot T_s = 0.83$ $\mathrm{ms}$. To reduce the localization overhead, the pilot signal is transmitted periodically every $d\tempo \gg \tau$. Localization and power allocation are thus performed every $d\tempo$ seconds, while the \ac{RIS} optimization procedure is performed every $T_O = N_{r}\, d\tempo$ seconds. We assume that the total available power is uniformly distributed over the entire spectrum, i.e., a power $P_{\text{tx}} = 0.06$ $\mathrm{mW}$ for the transmission of the pilot signal is allocated. We also consider a simple Rice channel model for the channel \ac{RIS}-\ac{UE} (see \eqref{eq:cascadedch2}), i.e., we assume that the \ac{LOS} component is always present, and we consider an uncorrelated \ac{NLOS} component.\footnote{The first assumption is reasonable because in the near-field region the distance between \ac{RIS} and \ac{UE} is small and the LOS probability is close to 1 \cite{3GPPloc}. As for the second assumption, uncorrelated scattering is a reasonable model in the absence of well-established channel models to characterize the \ac{RIS}-\ac{UE} channel.} 
\begin{figure}
    \centering
\centering
\input{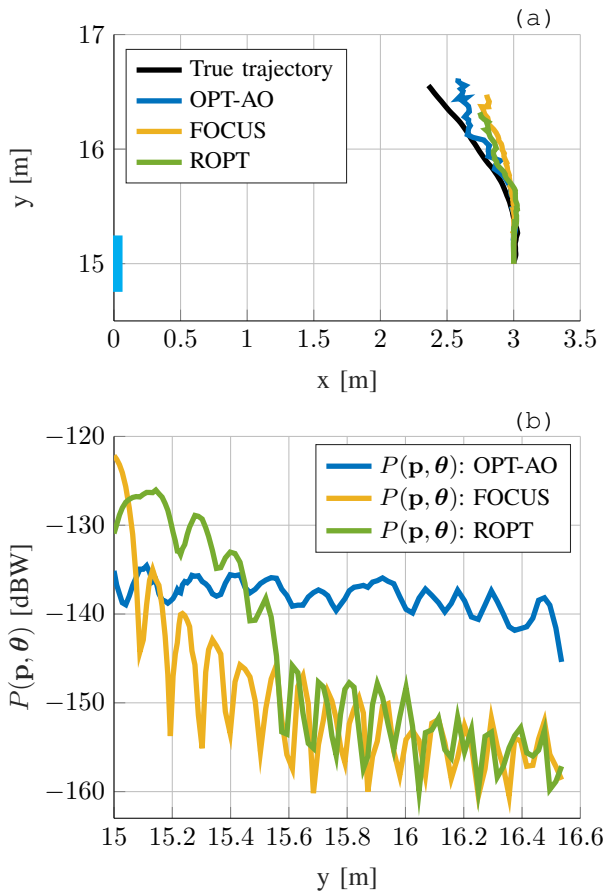}
    \caption{Comparison between OPT-AO, FOCUS, and ROPT for a 3-second trajectory in terms of: (a) estimated trajectory, (b) received power.
    }
    \label{fig:1RIS_1Traj}
\end{figure} 
\begin{figure*}
    \centering
\centering
\input{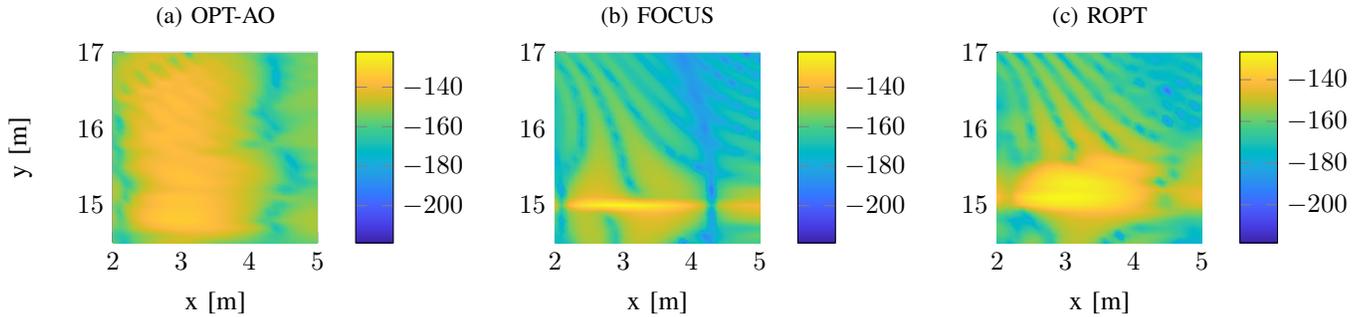}
\caption{{Distribution of reflected \ac{RIS} power in the scenario of Fig. \ref{fig:1RIS_1Traj} for the OPT-AO, FOCUS, and ROPT cases.}}
    \label{fig:1RIS_1Traj_map}
\end{figure*} 

The system geometry is sketched in Fig.~\ref{fig:scenario_sim}. The \ac{BS} is at a fixed position, namely at $\ptx=\left[30,\, 15,\, 2 \right]\,\mathrm{m}$, 
and it has a \ac{URA} with $8 \times 2$ antennas in the $YZ$-plane. 
The \ac{UE} changes positions through time according with \eqref{eq:motionmodel}\footnote{The $z$ coordinate is kept fixed at $1\,\mathrm{m}$.} and it is equipped with a \ac{ULA} of four antennas, lying horizontally along the $Y$-axis and with a fixed and known orientation. 
For the motion of the \ac{UE}, to emulate a movement pattern compatible with a typical random pedestrian motion model, we set $\sigma^2_{\mathsf{a},\mathsf{x}}=\sigma^2_{\mathsf{a},\mathsf{y}}= 0.5\, \mathrm{m}^2/\mathrm{s}^3$, 
$\sigma^2_{\mathsf{a},\mathsf{z}}= 0 \, \mathrm{m}^2/\mathrm{s}^3$. 
In all simulations, it is assumed that the initial state of the node (i.e., its position and velocity) is known, i.e. $\mathbf{m}_0=\mathbf{s}_0$ and $\boldsymbol{\Sigma}_0=\Q$. 
There are three \acp{RIS} in the environment located at $\left[0,\, 15,\, 3 \right]\,\mathrm{m}$,\, $\left[5,\, 0,\, 3 \right]\,\mathrm{m}$,\, $\left[10,\, 30,\, 3 \right]\,\mathrm{m}$, if not otherwise indicated. They consist of a \ac{URA} of $80 \times 5$ elements lying on the $YZ$-plane, i.e. they are mainly used in horizontal direction as long strips. In this way, the maximum dimension of the \ac{RIS} is nearly $40$ $\mathrm{cm}$ and the radiating near-field extends over about $32$ $\mathrm{m}$.

\paragraph{Tracking Scenario} 
In Fig. \ref{fig:1RIS_1Traj}(a), we consider a simplified scenario {to present a graphical representation of possible estimated trajectories and to show the corresponding RIS power distributions in Fig. \ref{fig:1RIS_1Traj}(b) and \ref{fig:1RIS_1Traj_map}. In this scenario } 
there is only the \ac{RIS} at position $(0,15)$ $\mathrm{m}$ and the terminal travels a 3-second trajectory starting at point $(3,15)$ with initial speed of $1$ $\mathrm{m/s}$ in the $y$ direction. \footnote{{
Since we are
considering a single RIS, all power is allocated to that RIS, and power allocation is not
effective, i.e., $\beta$OPT is equal to OPT.}}
The figure shows the actual node trajectory and the trajectories obtained considering \ac{RIS} optimization performed through the approaches OPT-AO, FOCUS, and ROPT. The \ac{RIS} is only optimized once at time $0$, while the location update  is performed every $d\tempo = 0.03$ seconds, i.e., we have $100$ location updates. For this short trajectory, having a single \ac{RIS} is sufficient to achieve good localization performance especially when the OPT-AO scheme is considered. In contrast, the ROPT and FOCUS optimization leads to larger estimation errors. 

The reason for this behavior can be explained by analyzing Fig. \ref{fig:1RIS_1Traj}(b), where the received power expressed in $\mathrm{dBW}$ is plotted as a function of the $y$ coordinate for the three systems under consideration. It is obvious that the FOCUS scheme maximizes the energy at the estimated point at time $0$ (i.e., when the \ac{UE} is at $y=15\,$ $\mathrm{m}$), but it leads to a huge power penalty as long as the terminal moves from its starting point. On the other hand, the OPT-AO scheme allows the power to be balanced over the entire interval, while the ROPT method concentrates the energy on the central part of the uncertainty range and spares the outer edges. Although this may be optimal for obtaining the maximum average rate, as shown below, it is inconvenient for localization.
 
In Fig. \ref{fig:1RIS_1Traj_map} we show the distribution of reflected \ac{RIS} power in the operational scenario of Fig. \ref{fig:1RIS_1Traj} for the cases OPT-AO, FOCUS, and ROPT, respectively. The power value at each point on the map is shown in false colors, with the color bar on the right indicating the power levels in $\mathrm{dBW}$. It can be seen that the OPT-AO approach{, Fig. \ref{fig:1RIS_1Traj_map}(a), } provides a nearly uniform distribution of the received power over the uncertainty range, the FOCUS approach{, Fig. \ref{fig:1RIS_1Traj_map}(b), } concentrates the power in a small area around the estimated endpoint $(3,15)$ $\mathrm{m}$, while the ROPT approach{, Fig. \ref{fig:1RIS_1Traj_map}(c), } provides uniform distribution in a limited area with respect to OPT-AO.  

\paragraph{Simulation Results} In the following, we report a comprehensive comparative simulation study considering the operating environment of Fig. \ref{fig:scenario_sim} with three \acp{RIS}. The \ac{UE} moves in a part of space with dimension $20$ $\mathrm{m}$ $\times$ $30$ $\mathrm{m}$ denoted as the range of motion (RM) corresponding to the yellow area in Fig. \ref{fig:scenario_sim}. Note that in the considered space, we are always in the near-field region for all considered \acp{RIS}. The \ac{UE} is initially placed randomly and moves along a 12-second trajectory according to the motion model considered in this work. Statistics are collected in $1000$ different simulations. 

\begin{figure}
    \centering
    \input{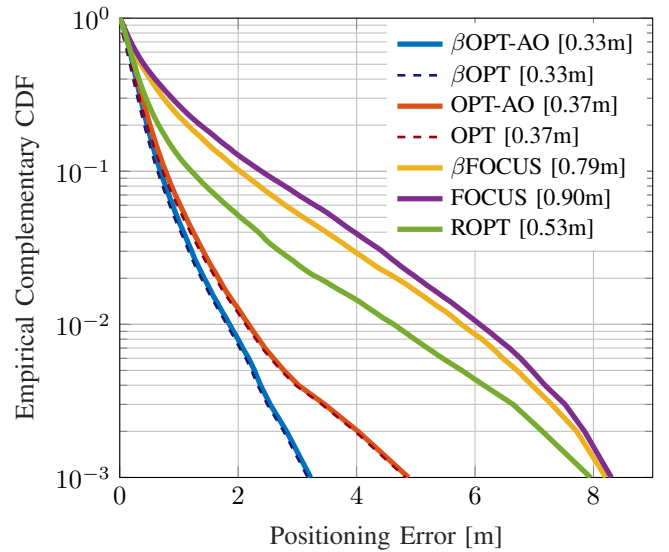}
    \caption{Empirical complementary \ac{CDF} of the localization error for different optimization approaches. In brackets, there are the mean values of the localization errors.} 
    \label{fig:con_e_senza_beta}
\end{figure}

Fig. \ref{fig:con_e_senza_beta} shows the empirical complementary \ac{CDF} \footnote{{The reason for reporting 1-cdf is to highlight the differences at low positioning
errors and to avoid possible curve overlapping.}} of the localization error, indicating the probability (rate) that the localization error is above a threshold, for the cases FOCUS, $\beta$FOCUS, OPT, $\beta$OPT, OPT-AO, $\beta$OPT-AO, and ROPT. The results are here obtained by considering a Rice factor $\kappa_b = 5$, $d\tempo = 0.03$ $\mathrm{s}$, and $N_{r} = 100${, i.e., $T_O = 3 \mathrm{s}$.  The choice of this RIS update time is connected with the specific motion model used for the UE. Indeed, by considering a typical random pedestrian motion model, as in \eqref{eq:motionmodel}, \eqref{eq:tmodel}, an update of $T_O=3 \,\mathrm{s}$ entails a UE movement of few meters that is compatible with the sub-meter localization accuracy. In the following, a further comparison for different $T_O$ values has been provided as well.}

First, we show that OPT-AO and $\beta$OPT-AO achieve the same performance as OPT and $\beta$OPT, confirming the validity of the AO approximation proposed in Sec.~\ref{RIsopt}. Moreover, the proposed approach OPT-AO significantly outperforms both FOCUS and ROPT, which confirms the validity of the proposed \ac{RIS} optimization for the localization procedure. Considering, in particular, $90\%$ \ac{CDF} (i.e., $10\%$ 1-\ac{CDF}), the results show that OPT-AO and $\beta$OPT-AO achieve an error below one meter. 
Regarding the mean localization errors (indicated in parentheses), we note that $\beta$OPT-AO has an average value of $33\, \mathrm{cm}$, which increases to $53$ and $79\, \mathrm{cm}$ for ROPT and $\beta$FOCUS, respectively. Finally, optimizing $\beta$ can lead to some improvement in localization error performance, which is particularly evident for the case OPT-AO at high \ac{CDF} values (i.e., at low 1-\ac{CDF}).

In Tab.~\ref{tab:rice_factors}, we provide the results for the case $\beta$OPT-AO for different values of the Rice factor $\kappa_b$. It can be seen that for the case shown in Fig. \ref{fig:con_e_senza_beta} ($\kappa_b = 5$), a slight performance degradation is observed compared to the LOS case ($\kappa_b = 100$). When the LOS component is further reduced, e.g., for $\kappa_b = 2$, the degradation increases as expected, although this does not affect the validity of the proposed approach.
\begin{table}
\begin{center}
\makebox[\linewidth]{
\begin{tabular}{ |c|c|c|c| } 
 \hline
 \rowcolor{silver} $\kappa_b$ & Mean value & 0.9 performance & 0.99 performance \\
  \hline
 $\kappa_b=2$ &0.38  & 0.78 & 2.17  \\ \hline
 $\kappa_b=5$ & 0.33  & 0.69 & 1.92 \\ \hline
 $\kappa_b=100$  & 0.29  & 0.59 & 1.58 \\ \hline
 \end{tabular}
}
\end{center}
\caption{Results for the case $\beta$OPT-AO for different values of the Rice factor $\kappa_b$ in terms of the localization error expressed in meters.}
\label{tab:rice_factors}
\end{table}

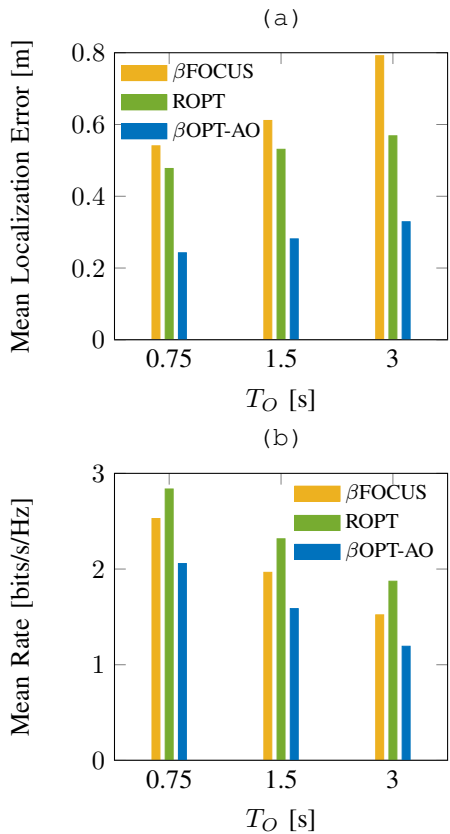
\begin{figure}[t!]
    \centering
%
%

\definecolor{mycolor1}{rgb}{0.00000,0.44700,0.74100}%
\definecolor{mycolor2}{rgb}{0.92900,0.69400,0.12500}%
\definecolor{mycolor3}{rgb}{0.46600,0.67400,0.18800}%
\begin{tikzpicture}

\begin{axis}[%
width=0.5\linewidth,
scale only axis,
bar shift auto,
xmin=0.511111111111111,
xmax=3.48888888888889,
xtick={1,2,3},
xticklabels={{0.75},{1.5},{3}},
ymin=0,
ymax=0.8,
ylabel= {Mean Localization Error [{m}]},
xlabel= {$T_O$ [s]},
axis background/.style={fill=white},
title style={font=\ttfamily},
title={(a)},
legend style={legend cell align=left, align=left, draw=white!15!black,at={(0.48,1)}, font=\footnotesize,fill=none,draw=none}
]
\addplot[ybar, bar width=3, draw=mycolor2,fill=mycolor2, area legend] table[row sep=crcr] {%
1	0.540542464000624\\
2	0.611187239533506\\
3	0.791696134927278\\
};
\addplot[forget plot, color=white!15!black] table[row sep=crcr] {%
0.511111111111111	0\\
3.48888888888889	0\\
};
\addlegendentry{$\beta\text{FOCUS}$}

\addplot[ybar, bar width=3, draw=mycolor3,fill=mycolor3, area legend] table[row sep=crcr] {%
1	0.477384258713429\\
3	0.568451403187324\\
2	0.530862492976419\\
};
\addplot[forget plot, color=white!15!black] table[row sep=crcr] {%
0.511111111111111	0\\
3.48888888888889	0\\
};
\addlegendentry{ROPT}

\addplot[ybar, bar width=3, draw=mycolor1, fill=mycolor1, area legend] table[row sep=crcr] {%
1	0.242465487847912\\
2	0.281228366627421\\
3	0.328924068053089\\
};
\addplot[forget plot, color=white!15!black] table[row sep=crcr] {%
0.511111111111111	0\\
3.48888888888889	0\\
};
\addlegendentry{$\beta\text{OPT-AO}$}

\end{axis}
\end{tikzpicture}%
%
%

\definecolor{mycolor1}{rgb}{0.00000,0.44700,0.74100}%
\definecolor{mycolor2}{rgb}{0.92900,0.69400,0.12500}%
\definecolor{mycolor3}{rgb}{0.46600,0.67400,0.18800}%
\begin{tikzpicture}

\begin{axis}[%
width=0.5\linewidth,
scale only axis,
bar shift auto,
xmin=0.511111111111111,
xmax=3.48888888888889,
xtick={1,2,3},
xticklabels={{0.75},{1.5},{3}},
ymin=0,
ymax=3,
axis background/.style={fill=white},
title style={font=\ttfamily},
title={(b)},
ylabel={Mean Rate [{bits/s/Hz}]},
xlabel={$T_O$ [s]},
legend style={legend cell align=left, align=left, draw=white!15!black, font=\footnotesize, draw=none, fill=none, at={(1,1)}}
]
\addplot[ybar, bar width=3, draw=mycolor2,fill=mycolor2, area legend] table[row sep=crcr] {%
1	2.52759731888125\\
2	1.96558711334276\\
3	1.52122965487485\\
};
\addplot[forget plot, color=white!15!black] table[row sep=crcr] {%
0.511111111111111	0\\
3.48888888888889	0\\
};
\addlegendentry{$\beta\text{FOCUS}$}

\addplot[ybar, bar width=3, draw=mycolor3,fill=mycolor3, area legend] table[row sep=crcr] {%
1	2.83653641425722\\
2	2.31589524930695\\
3	1.87263968063888\\
};
\addplot[forget plot, color=white!15!black] table[row sep=crcr] {%
0.511111111111111	0\\
3.48888888888889	0\\
};
\addlegendentry{ROPT}

\addplot[ybar, bar width=3, draw=mycolor1,fill=mycolor1, area legend] table[row sep=crcr] {%
1	2.05719942362025\\
2	1.58636792331726\\
3	1.19303834680262\\
};
\addplot[forget plot, color=white!15!black] table[row sep=crcr] {%
0.511111111111111	0\\
3.48888888888889	0\\
};
\addlegendentry{$\beta\text{OPT-AO}$}

\end{axis}

\end{tikzpicture}%
    \caption{{Mean localization error (top) and mean rate (bottom) for $\beta$FOCUS, $\beta$OPT-AO, and ROPT algorithms computed for different $T_O$ values and $d\tempo=0.03$ seconds.}}
    \label{fig:different_Time_opt_histogram_error}
\end{figure}

{We also provide some results on communication rates and localization errors, which depend on how the objective function for the RIS optimization problem is defined.
In our work, the RIS optimization problem is formulated such that the
UE tracking performance is maximized (e.g., solved by the $\beta$OPT-AO algorithm). Therefore, we expect the results to be better in terms of positioning accuracy than if RIS had been optimized in favor of communication and vice versa. To better
illustrate this point, we refer to }
Fig. \ref{fig:different_Time_opt_histogram_error} {in which } the mean localization error (top) and the average achievable rate (bottom) for the algorithms $\beta$FOCUS, $\beta$OPT-AO and ROPT are shown for different $T_O$ values and for $d\tempo = 0.03$ $\mathrm{s}$. As expected, the localization error decreases with the decrease of $T_O$ for all methods considered, while $\beta$OPT-AO clearly outperforms the other methods in all cases. {Regarding the communication part, we do not consider additional pilots that would be necessary for the CSI estimation and the corresponding precoder optimization, as these aspects are beyond the scope of this work. Instead, we consider Shannon's standard formula to obtain an estimate of the achievable rate. In other words} it is defined as the capacity in $\mathrm{bits/s/Hz}$ of the channel matrix $\mathbf{H}_{eq} = \sum_{k=1}^{K} \mathbf{B}_k \RISPARAMk \mathbf{G}_k$ with noise variance $\sigma^2$, which can be determined by \ac{SVD} and waterfilling.
 Interestingly, it can be observed that the $\beta$OPT-AO approach, which is the best in terms of localization error, is the worst in terms of mean achievable rate. Moreover, the highest mean achievable rate is obtained by ROPT, as expected.
{A more detailed consideration can be made by comparing the algorithms $\beta$OPT-AO and ROPT and considering the case $T_O = 3\, \mathrm{s}$, which shows that the proposed method $\beta$OPT-AO achieves a rate almost 35\% lower than that of ROPT, which was specifically designed for communication purposes. Conversely, it achieves 42\% lower positioning error than ROPT. The same happens in all other analyzed cases for different $T_O$ and $d\mathrm{t}$ values, so a trade-off between localization and communication can be found. This is a non-obvious result, since it is generally assumed that the best RIS optimization strategy is always to maximize the power in the direction where the node is (or should be) located.}

%
\begin{figure}[t!]
    \centering
    \input{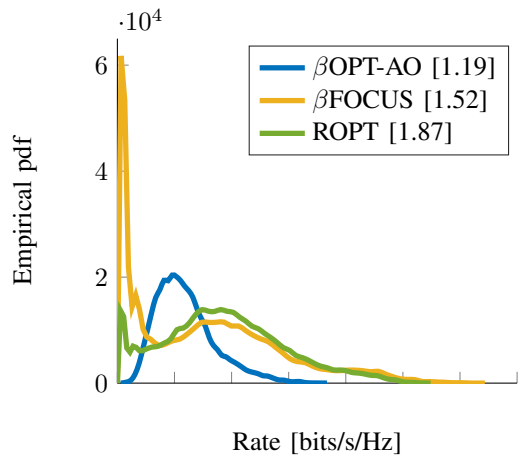}
    \caption{Empirical pdf of the achievable average rate for $\beta$FOCUS, $\beta$OPT-AO, and ROPT.}
    \label{fig:hist_rate}
\end{figure} 

In Fig. \ref{fig:hist_rate}, we give the empirical \ac{pdf} of the achievable rate for $\beta$FOCUS, $\beta$OPT-AO, and ROPT. It can be seen that the $\beta$FOCUS and $\beta$OPT-AO approaches, although better than $\beta$OPT-AO in terms of mean rate, have high probabilities of zero or quasi-zero rate, while $\beta$OPT-AO avoids such unpleasant situations.
\begin{figure}[t!]
    \centering
%
%

\definecolor{mycolor1}{rgb}{0.00000,0.44700,0.74100}%
\definecolor{mycolor2}{rgb}{0.92900,0.69400,0.12500}%
\definecolor{mycolor3}{rgb}{0.46600,0.67400,0.18800}%

\begin{tikzpicture}

\begin{axis}[%
width=0.5\linewidth,
scale only axis,
bar shift auto,
xmin=0.511111111111111,
xmax=3.48888888888889,
xtick={1,2,3},
xticklabels={{0.01},{0.03},{0.1}},
ymin=0,
ymax=1.2,
xlabel= {$\red{d\mathsf{t}}$ [s]},
axis background/.style={fill=white},
title style={font=\ttfamily},
title={(a)},
ylabel={Mean Localization Error [{m}]},
legend style={legend cell align=left, align=left, draw=white!15!black, at={(0.49,0.99)}, font=\footnotesize, draw=none, fill=none}
]
\addplot[ybar, bar width=3, draw=mycolor2,fill=mycolor2, area legend] table[row sep=crcr] {%
1	0.509498069466471\\
3	1.17713733779858\\
2	0.791696134927278\\
};
\addplot[forget plot, color=white!15!black] table[row sep=crcr] {%
0.511111111111111	0\\
3.48888888888889	0\\
};
\addlegendentry{$\beta\text{FOCUS}$}

\addplot[ybar, bar width=3, draw=mycolor3,fill=mycolor3, area legend] table[row sep=crcr] {%
1	0.380335938150045\\
3	0.870639322464108\\
2	0.530862492976419\\
};
\addplot[forget plot, color=white!15!black] table[row sep=crcr] {%
0.511111111111111	0\\
3.48888888888889	0\\
};
\addlegendentry{ROPT}

\addplot[ybar, bar width=3, draw=mycolor1,fill=mycolor1, area legend] table[row sep=crcr] {%
1	0.208979009431203\\
3	0.465472650076725\\
2	0.328924068053089\\
};
\addplot[forget plot, color=white!15!black] table[row sep=crcr] {%
0.511111111111111	0\\
3.48888888888889	0\\
};
\addlegendentry{$\beta\text{OPT-AO}$}

\end{axis}

\end{tikzpicture}%
%
%

\definecolor{mycolor1}{rgb}{0.00000,0.44700,0.74100}%
\definecolor{mycolor2}{rgb}{0.92900,0.69400,0.12500}%
\definecolor{mycolor3}{rgb}{0.46600,0.67400,0.18800}%

\begin{tikzpicture}

\begin{axis}[%
width=0.5\linewidth,
scale only axis,
bar shift auto,
xmin=0.511111111111111,
xmax=3.48888888888889,
xtick={1,2,3},
xticklabels={{0.01},{0.03},{0.1}},
ymin=0,
ymax=2.5,
xlabel= {$\red{d\mathsf{t}}$ [s]},
axis background/.style={fill=white},
title style={font=\ttfamily},
title={(b)},
ylabel={Mean Rate [bits/s/Hz]},
legend style={legend cell align=left, align=left, draw=white!15!black, at={(1,1.01)}, font=\footnotesize, draw=none, fill=none}
]
\addplot[ybar, bar width=3, draw=mycolor2,fill=mycolor2, area legend] table[row sep=crcr] {%
1	1.69451643124713\\
3	1.42855900729268\\
2	1.52122965487485\\
};
\addplot[forget plot, color=white!15!black] table[row sep=crcr] {%
0.511111111111111	0\\
3.48888888888889	0\\
};
\addlegendentry{$\beta\text{FOCUS}$}

\addplot[ybar, bar width=3, draw=mycolor3,fill=mycolor3, area legend] table[row sep=crcr] {%
1	2.0778932609222\\
3	1.74083635081085\\
2	1.87263968063888\\
};
\addplot[forget plot, color=white!15!black] table[row sep=crcr] {%
0.511111111111111	0\\
3.48888888888889	0\\
};
\addlegendentry{ROPT}

\addplot[ybar, bar width=3, draw=mycolor1,fill=mycolor1, area legend] table[row sep=crcr] {%
1	1.24892015120383\\
3	1.12279644695086\\
2	1.19303834680262\\
};
\addplot[forget plot, color=white!15!black] table[row sep=crcr] {%
0.511111111111111	0\\
3.48888888888889	0\\
};
\addlegendentry{$\beta\text{OPT-AO}$}

\end{axis}

\end{tikzpicture}%
    \caption{{Mean localization error (top) and mean rate (bottom) for $\beta$FOCUS, $\beta$OPT-AO, and ROPT algorithms computed for different $d\tempo$ values and $T_O=3$ seconds.}}
    \label{fig:different_dt_histogram_rate}
\end{figure}
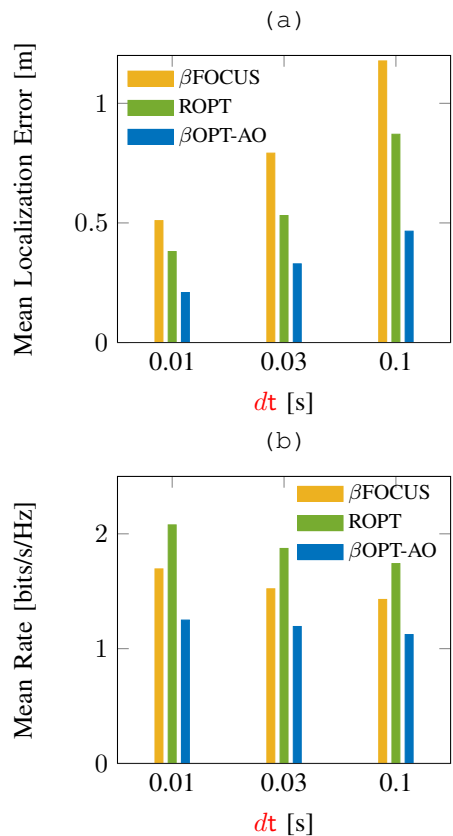 

In Fig. \ref{fig:different_dt_histogram_rate}, the results for different $d\tempo$ values (in seconds) and for $T_O = 3\, \mathrm{s}$ are shown for the $\beta$FOCUS, $\beta$OPT-AO, and ROPT algorithms. As expected the mean localization error{, Fig. \ref{fig:different_dt_histogram_rate}(a), } decreases with the decrease of $d\tempo$ for all considered methods while the achievable rate{, Fig. \ref{fig:different_dt_histogram_rate}(b), } increases correspondingly. The opposite behavior is observed if $d\tempo$ increases.

\begin{figure}
    \centering
    \input{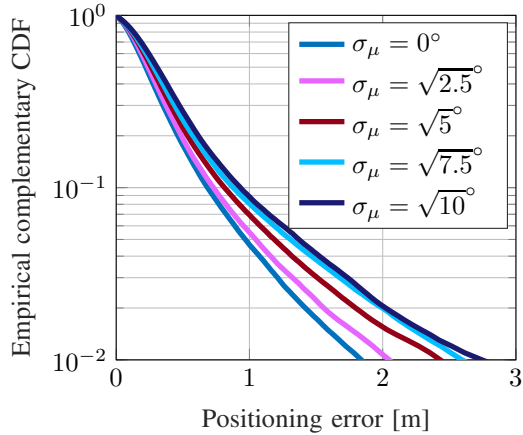}
    \caption{{Empirical complementary cdf of the positioning error for different orientation errors $ \epsilon_{\mu}\sim \mathcal{N}(0,\sigma_{\mu}^2)$ with $\epsilon_{\mu}=\mu- \hat{\mu}$ being the orientation error computed as the difference of the true orientation angle ($\mu$) and its estimate ($\hat{\mu}$), and with $\sigma_{\mu}$ being the standard deviation of the orientation estimator in degrees ([$^\circ$]).}}
    \label{fig:rot_error}
\end{figure}

{In Fig. \ref{fig:rot_error} we report the complementary empirical cdf of the positioning error for different orientation errors. This analysis allows testing the robustness of the proposed $\beta$OPT-AO in the presence of a residual estimation error when the \ac{UE} orientation is inferred from inertial devices (e.g., IMU) \cite{Ngu22, Gao23}. Therefore, we performed simulations in which the orientation of the UE is affected by an i.i.d. unbiased Gaussian error with a standard deviation between $\sigma_{\mu}=1^\circ$ and $\sigma_{\mu}=\sqrt{10}^\circ$. The result is promising as it shows that in all cases the localization error is below $1\,$m for $90\%$ of the time.}

\begin{figure}[t!]
    \centering
    \input{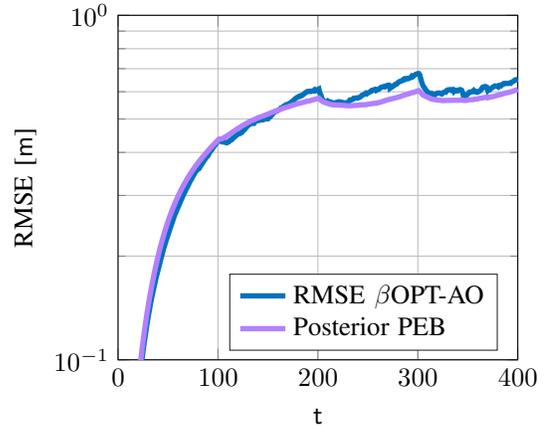}
    \caption{{
    Logarithmic representation of {the p}osterior \ac{PEB} vs \ac{RMSE} for the $\beta$OPT-AO algorithm.}}
    \label{fig:PEB}
\end{figure}

{Finally, in Fig. \ref{fig:PEB} we show the RMSE performance of our proposed algorithm ($\beta$OPT-AO) compared with the \ac{PEB} \cite{tichavsky1998posterior}. The results show a wavelike behavior corresponding to the RIS optimization {instants} (one every $100$ iterations) and a perfect match between the bound and the algorithm. The figure also shows the transition time (of almost $150$ iterations) where the error increases, since we assumed an initial time with zero error.
}

\section{Conclusions}
\label{sec:conclusions}
In this paper, we have proposed a novel framework to jointly design the reflection coefficients of multiple \acp{RIS} and the precoding strategy of a single \ac{BS}, with two different time scales, to optimize the tracking of a single multi-antenna \ac{UE} and to keep the overall system complexity affordable. The optimal derived \ac{RIS} and precoding strategy has been compared with focusing and  rate maximization, showing that  \ac{RIS} optimization for communication is suboptimal when used for tracking purposes.
{Numerical results have shown the possibility of achieving high-accuracy positioning performance in typical indoor environments with a single \ac{BS} and a few \acp{RIS} at millimeter-waves. This is in accordance with the trend and requirements foreseen for the next 6G networks.} 
Simulation results have revealed that having multiple and bigger \acp{RIS} can lead to a significant increase of the tracking performance provided that are properly jointly optimized.

\section*{Appendix A: Jacobian Computation}

The first derivatives of the observation functions can be written as
\begin{align}\label{eq:der1_v2}
&\frac{\partial \mathbb{E}\left\{\Re\left\{\rhorrk \right\} \right\}}{\partial \varsigma}= \nonumber\\
 &\Reakrr \frac{\partial \Reakrrp }{\partial \varsigma} + \Reakrrp \frac{\partial \Reakrr }{\partial \varsigma} \nonumber\\
 &+ \Imakrr \frac{\partial \Imakrrp }{\partial \varsigma} + \Imakrrp \frac{\partial \Imakrr }{\partial \varsigma},
\end{align}
where $\varsigma \in \left\{x,\,y,\,z \right\}$ is a generic position coordinate. A similar calculation applies to the imaginary parts.
According to the received signal, we have
\begin{align}
\label{eq:der3}
 \frac{\partial \nark}{\partial \varsigma} & \approx \frac{1}{\lvert a_{\risk} \rvert} \frac{\partial \ark }{\partial \varsigma},
\end{align}
where we assumed that the amplitude of the received signal changes slowly with respect to the phase and where
\begin{align}\label{eq:der4}
 \frac{\partial \ark }{\partial \varsigma}= & - \jmath \frac{2\pi}{\lambda} \sqrt{\frac{\kappa_{b}}{\kappa_{b}+1}} \sum_{\tx=0}^{\Ntx-1} \ftk \sum_{\riselem=0}^{P-1} \rhokpr\,\rhoktp\, \, \nonumber \\
 &\times \tkp \,
 e^{- \jmath \frac{2\pi}{\lambda}\, d_{\risk, \rx,\riselem,\tx} } \frac{\partial d_{\risk; \rx,\riselem,\tx}}{\partial \varsigma}.
\end{align}
 To simplify the notation $d_{\risk, \rx,\riselem,\tx} \triangleq \dktp + \dkpr$. From \eqref{eq:der4}, the derivatives of $\Re\left\{ \nark \right\}$ and $\Im\left\{ \nark \right\}$ can be easily derived by taking into account that
\begin{align}
&\frac{\partial d_{\risk; \rx,\riselem,\tx}}{\partial x}=\frac{\partial \dkpr}{\partial x} = \frac{x_{\risk, \riselem}-x_{\rx}}{\dkpr}, \\
&\frac{\partial d_{\risk; \rx,\riselem,\tx}}{\partial y}=\frac{\partial \dkpr}{\partial y}=  \frac{y_{\risk, \riselem}-y_{\rx}}{\dkpr}, \\
&\frac{\partial d_{\risk; \rx,\riselem,\tx}}{\partial z} =\frac{\partial \dkpr}{\partial z}=  \frac{z_{\risk, \riselem}-z_{\rx}}{\dkpr},
\end{align}
and where we assume that the orientations of all devices are known.
The derivatives with respect to the velocity are zeros, since the observations refer only to the position of \ac{UE}. Therefore, the velocity is derived using only the transition model and the latest position estimates (without measurement correction/update, which applies only to the position estimate).

\section*{Appendix B}
In this Appendix we show how the problem expressed in
\begin{equation}\label{eq:problem111}
\begin{array}{c}
 \underset{\substack{{\txparam^{(\tempop)}}}} 
 {\min} ~~ \mathbb{E}_{{\sskp \lvert \tempo}} \left\{\mathcal{E}\left({\mathbf{p}_{\tempop},{\txparam^{(\tempop)}}}\right)\right\} \\
\text{s.t.}
\quad {\txparam^{(\tempop)}} \in {\mathcal{Q}} 
\end{array}
\end{equation}
can be recasted as
 \begin{equation}\label{eq:problem1111}
\begin{array}{c}
\underset{\substack{{\txparam^{(\tempop)}}}} 
 {\min} ~~ \text{Tr} \left({\tilde{\mathbf{K}}_{\tempop}}  ~ \mathbb{E}_{{\sskp \lvert \tempo }} \left\{\Rkp\left({\mathbf{p}_{\tempop}, {\txparam^{(\tempop)}}}\right)\right\} {\tilde{\mathbf{K}}_{\tempop}}^T \right)\\
\text{s.t.}
\quad {{\txparam^{({\tempop})}}} \in {\mathcal{Q}}  \, 
\end{array} 
\end{equation} 

By considering that the cost function can be written as 
\begin{align}
\label{eq:costfunctionwy}
&\mathcal{E}\left({\mathbf{p}_{\tempop},{\txparam^{(\tempop)}}}\right) = \text{Tr} (\Delta\sskp\Delta\sskp^T  
\nonumber \\
 & 
 + {\tilde{\mathbf{K}}_{\tempop}} \left[ \Hkp\,\Delta\sskp\Delta\sskp^T\,\Hkp^T + \Rkp\right] {\tilde{\mathbf{K}}_{\tempop}}^T  \nonumber \\
 & - {\tilde{\mathbf{K}}_{\tempop}}\Hkp \Delta\sskp\Delta\sskp^T  - \Delta\sskp\Delta\sskp^T \Hkp^T {\tilde{\mathbf{K}}_{\tempop}}^T  ) ,
\end{align}
where we have used the approximation ${\mathbf{K}}_{\tempop} \approx {\tilde{\mathbf{K}}_{\tempop}}$, we can apply the expectation, namely {$\mathbb{E}_{\sskp \lvert \tempo}=\mathbb{E}$}, as
\begin{align}
\label{eq:costfunctionwy_der}
&\mathbb{E}\left\{\mathcal{E}\left({\mathbf{p}_{\tempop},{\txparam^{(\tempop)}}}\right) \right\} =\text{Tr} ( \mathbb{E}\left\{\Delta\sskp\Delta\sskp^T \right\} 
\nonumber \\
 & 
 + {\tilde{\mathbf{K}}_{\tempop}} \left[ \Hkp\,\mathbb{E}\left\{\Delta\sskp\Delta\sskp^T \right\}\,\Hkp^T + \mathbb{E}\left\{\Rkp \right\}\right] {\tilde{\mathbf{K}}_{\tempop}}^T  \nonumber \\
 & - {\tilde{\mathbf{K}}_{\tempop}}\Hkp \mathbb{E}\left\{\Delta\sskp\Delta\sskp^T \right\} \nonumber \\
 &- \mathbb{E}\left\{\Delta\sskp\Delta\sskp^T \right\}\Hkp^T {\tilde{\mathbf{K}}_{\tempop}}^T  ) \nonumber \\
 &=\text{Tr} ( \boldsymbol{\Sigma}_{\tempop \lvert \tempo}
+ {\tilde{\mathbf{K}}_{\tempop}} \left[ \Hkp\,\boldsymbol{\Sigma}_{\tempop \lvert \tempo}\,\Hkp^T + \mathbb{E}\left\{\Rkp\right\}\right] {\tilde{\mathbf{K}}_{\tempop}}^T  \nonumber \\
 & - {\tilde{\mathbf{K}}_{\tempop}}\Hkp \boldsymbol{\Sigma}_{\tempop \lvert \tempo}- \boldsymbol{\Sigma}_{\tempop \lvert \tempo}\Hkp^T {\tilde{\mathbf{K}}_{\tempop}}^T  ). 
\end{align}
In \eqref{eq:costfunctionwy_der}
 the only term that depends on $\txparam^{(\tempop)}$ is 
 $\Rkp$. Therefore we have
\begin{align}
\label{eq:costfunctionwy_der1}
&\mathbb{E}\left\{\mathcal{E}\left({\mathbf{p}_{\tempop},{\txparam^{(\tempop)}}}\right) \right\} = \text{Tr} ( {\tilde{\mathbf{K}}_{\tempop}} \, \mathbb{E}\left\{\Rkp\right\}\, {\tilde{\mathbf{K}}_{\tempop}}^T).
\end{align}
Finally it can be noted that $\Rkp$ only depends on the \ac{UE} position, hence $\mathbb{E}_{{\sskp \lvert \tempo}}\left\{\Rkp\right\}=\mathbb{E}_{{\mathbf{p}_{\tempop} \lvert {\tempo} }}\left\{\Rkp\right\}$.

\bibliographystyle{IEEEtran}


\end{document}